\documentstyle[12pt,psfig,axodraw]{article}
\newlength{\dinwidth}
\newlength{\dinmargin}
\mathchardef\less=316
\mathchardef\greater=318
\def\gev{{\rm Ge}\kern-1.pt{\rm V}}
\def\3{\ss}
\newcommand{\ptmiss}{{\mbox{$\not\hspace{-.55ex}{P}_t^{cal}$}}}
\newcommand{\rpvio}{{\not\hspace{-.55ex}{R}_P}}
\newcommand{\noe}{{\mbox{$\not\hspace{-.5ex}e$}}}

\newcommand{\shat}{\mbox{${\hat s}$}}
\newcommand{\sTop}{{\tilde t}}

\newcommand{\photino}{{\tilde\gamma}}
\newcommand{\MLQ}{\mbox{$M_{\scriptscriptstyle LQ}$}}
\newcommand{\MLQsq}{\mbox{$M_{\scriptscriptstyle LQ}^2$}}
\newcommand{\GammaLQ}{\mbox{$\Gamma_{\scriptscriptstyle LQ}$}}
\newcommand{\GammaLQsq}{\mbox{$\Gamma_{\scriptscriptstyle LQ}^2$}}
\newcommand{\MLQr}{\mbox{$\mu_{\scriptscriptstyle LQ}^{\scriptscriptstyle rec}$}}
\newcommand{\sigLQr}{\mbox{$\sigma_{\scriptscriptstyle LQ}^{\scriptscriptstyle rec}$}}
\newcommand{\invis}{^{\scriptscriptstyle invis}}
\newcommand{\miss}{\mbox{$_{\scriptscriptstyle miss}$}}
\newcommand{\clu}{\mbox{$_{\scriptscriptstyle clu}$}}
\newcommand{\phiclubar}{\mbox{${\overline{\phiclu}}$}}
\newcommand{\phiclu}{\phi\clu}

\newcommand{\lum}{\mbox{${\cal L}$}}
\newcommand{\ipb}{\mbox{$pb^{-1}$}\,}
\newcommand{\tV}{\tilde V}
\newcommand{\tS}{\tilde S}
\newcommand{\lepton}{{\ell}}
\newcommand{\ptl}{{\mbox{${P}_t^\lepton$}}}
\begin{document}
\vspace{1 cm}
\begin{titlepage}

\title{\bf Search for Lepton Flavor Violation \\
           in $ep$ Collisions at 300 GeV \\
          Center of Mass Energy}

\author{ZEUS Collaboration}
\date{}
\maketitle

\vspace{5 cm}

\begin{abstract}
\centerline{\vbox{\hsize14.cm
\noindent
Using the ZEUS detector at the HERA electron-proton collider, we have 
searched for lepton flavor violation in $ep$ collisions at a
center--of--mass energy ($\sqrt{s}$) of 300 GeV.
Events of the type $e+p\to \lepton +X$ with a final--state lepton of high
transverse momentum, $\lepton=\mu$ or $\tau$, were sought. No evidence was found
for lepton flavor violation in the combined 1993 and 1994 data samples, for
which the integrated luminosities were 0.84 \ipb for $e^-p$
collisions and 2.94 \ipb for $e^+p$ collisions.  Limits on
coupling {\it vs.}~mass are provided for leptoquarks and
$R$--parity violating squarks.  For flavor violating
couplings of electromagnetic strength, we set 95\% confidence level lower limits
on leptoquark masses between 207 GeV and 272 GeV, depending on the leptoquark
species and final--state lepton. For leptoquark masses larger than
300 GeV, limits on flavor--changing couplings are determined, many of which
supersede prior limits from rare decay processes.
}}
\end{abstract}
\begin{center}
To be published in Zeitschrift f\"ur Physik C.
\end{center}
\vskip -21cm
\centerline{{\tt DESY 96-161}\hfill{\tt ISSN 0418-9833}}
\setcounter{page}{0}
\thispagestyle{empty}   
\end{titlepage}
\begin{center}
{                      \Large  The ZEUS Collaboration              }
\end{center}
  M.~Derrick,
  D.~Krakauer,
  S.~Magill,
  D.~Mikunas,
  B.~Musgrave,
  J.R.~Okrasi\'{n}ski,
  J.~Repond,
  R.~Stanek,
  R.L.~Talaga,
  H.~Zhang  \\
 {\it Argonne National Laboratory, Argonne, IL, USA}~$^{p}$
% ...
\newpage
% ...
$^{  19}$ now at ATLAS Collaboration, Univ. of Munich \\
% ...
                                                           %
                                                           %
\newpage   
                                                           %
                                                           %
\begin{tabular}[h]{rp{14cm}}
$^{a}$ &  supported by the Natural Sciences and Engineering Research
          Council of Canada (NSERC)  \\
$^{b}$ &  supported by the FCAR of Qu\'ebec, Canada  \\
$^{c}$ &  supported by the German Federal Ministry for Education and
          Science, Research and Technology (BMBF), under contract
          numbers 057BN19P, 057FR19P, 057HH19P, 057HH29P, 057SI75I \\
$^{d}$ &  supported by the MINERVA Gesellschaft f\"ur Forschung GmbH,
          the Israel Academy of Science and the U.S.-Israel Binational
          Science Foundation \\
$^{e}$ &  supported by the German Israeli Foundation, and
          by the Israel Academy of Science  \\
$^{f}$ &  supported by the Italian National Institute for Nuclear Physics
          (INFN) \\
$^{g}$ &  supported by the Japanese Ministry of Education, Science and
          Culture (the Monbusho) and its grants for Scientific Research \\
$^{h}$ &  supported by the Korean Ministry of Education and Korea Science
          and Engineering Foundation  \\
$^{i}$ &  supported by the Netherlands Foundation for Research on
          Matter (FOM) \\
$^{j}$ &  supported by the Polish State Committee for Scientific
          Research, grants No.~115/E-343/SPUB/P03/109/95, 2P03B 244
          08p02, p03, p04 and p05, and the Foundation for Polish-German
          Collaboration (proj. No. 506/92) \\
$^{k}$ &  supported by the Polish State Committee for Scientific
          Research (grant No. 2 P03B 083 08) and Foundation for
          Polish-German Collaboration  \\
$^{l}$ &  partially supported by the German Federal Ministry for
          Education and Science, Research and Technology (BMBF)  \\
$^{m}$ &  supported by the German Federal Ministry for Education and
          Science, Research and Technology (BMBF), and the Fund of
          Fundamental Research of Russian Ministry of Science and
          Education and by INTAS-Grant No. 93-63 \\
$^{n}$ &  supported by the Spanish Ministry of Education
          and Science through funds provided by CICYT \\
$^{o}$ &  supported by the Particle Physics and
          Astronomy Research Council \\
$^{p}$ &  supported by the US Department of Energy \\
$^{q}$ &  supported by the US National Science Foundation \\
\end{tabular}
                                                           %
                                                           %
\section{Introduction}\label{intro}

Lepton flavor is conserved in all interactions of the Standard Model~(SM);
discovery of lepton--flavor violation (LFV) in any form would be evidence
for physics beyond our principal particle physics paradigm.  Many searches
for specific reactions which violate lepton flavor have been performed.
The most sensitive include searches
for $\mu +N\to e +N$ using very low--energy muons\cite{Tiexp},
for the forbidden muon decay $\mu\to e\gamma$ \cite{muegam},
and for forbidden leptonic kaon decays\cite{rareK}.
The limits from these processes are sensitive to  
$e\leftrightarrow\mu$ flavor change, but not to $e\leftrightarrow\tau$.
Also, each of these processes involves specific quark flavors:
in the first case, only first generation quarks participate; in the second
case, for mechanisms which involve virtual quarks,
the same quark flavor must couple to both $e$ and $\mu$; in the last
case, strange quarks must be involved.  Since lepton flavor change could
involve the $\tau$-lepton or could be
accompanied by quark flavor change, there may be LFV reactions which
would be invisible to these very sensitive experiments.
Hence, we have no {\it a priori} reason to assume that flavor
violation, should it exist, will be visible through these specific
reactions. Therefore, though less
sensitive in an absolute sense, other manifestations of flavor violation,
like forbidden leptonic decays of $B$-- and $D$--mesons and of
$\tau$--leptons, are being investigated\,\cite{BDdecay}.

We report here a search for LFV carried out by the ZEUS collaboration
at the HERA $ep$ collider where we have sought instances of the reaction 
\begin{equation}
e+p \rightarrow \lepton+X  \label{epreac},
\end{equation}
where $\lepton$ represents an isolated final--state $\mu $ or $\tau $ with large
transverse momentum and $X$ represents the hadronic final state.  
Processes with such topologies can be found in ZEUS
with good efficiency and with little background. It should be emphasized
that any reaction of the type (\ref{epreac}) in which a final--state
high--energy $\mu$ or $\tau$ replaces the incident electron\footnote{In the
following, ``electron" is
generically used to denote both electrons and positrons.}
would be direct evidence for physics beyond 
the Standard Model, independent of the underlying mechanism.
Furthermore, this reaction
should occur at some level for a wide range of possible LFV mechanisms.
LFV\ mechanisms which also involve a quark flavor change,
or which are stronger for heavier quarks \cite{gclq}
may be seen more readily at HERA, where the sensitivity is largely
independent of quark flavor\footnote{
Though the threshold for top ($t$) quark production is below the HERA
center--of--mass energy, with present luminosities $t$ production would
only be observable if the couplings were very large. Therefore, we choose
not to report on LFV couplings involving top in this paper.},
than in low--energy experiments. 

The lepton--flavor violating reaction~(\ref{epreac}) could
occur via $s$--, $u$--, or $t$--channel
exchanges as shown in figure ~\ref{fig:stugraph}. For
the $s$-- and $u$--channel processes the exchanged particle has
the quantum numbers of a leptoquark (or an $R$--parity violating squark).  The
cross sections depend on the leptoquark species and mass, and on the
couplings, $\lambda_{eq_1}$ and $\lambda_{\lepton q_2}$ shown in
figure \ref{fig:stugraph}.
For the case of $t$--channel exchange, the process would be mediated
by a flavor--changing neutral boson.

For definiteness, we will describe reaction (\ref{epreac}) with
leptoquarks as the carrier of the LFV force,
treating separately the cases of direct leptoquark production and the
virtual effects of leptoquarks with masses above 300 GeV.
The similarity of production formulae between $R$--parity--violating squarks
and certain leptoquarks permits us to relate the couplings
implied by the two mechanisms for a specified cross section.
Results on flavor violation induced by leptogluons or
by flavor--changing neutral bosons as well as details of the technique
used in this analysis are also available \cite{songhoon}.
The H1 collaboration \cite{H1} has also searched for direct production
of leptoquarks with flavor--violating couplings using similar methods.

This analysis is based on an integrated luminosity of 0.84 \ipb
(2.94 \ipb) of $e^-p$ ($e^+p$) data taken during the 1993 and
1994 running periods. Beam energies at HERA were 26.7 GeV
(27.5 GeV) for the electron beam in 1993
(1994) and 820 GeV for the proton beam. The resulting center--of--mass
energy of 296 GeV (300 GeV) is an order of magnitude higher
than for fixed--target lepton--nucleon scattering experiments.

\section{Scenarios of Lepton Flavor Violation}\label{LFVsce}
\subsection{Leptoquarks}\label{LQsce}

A leptoquark (LQ) is a hypothetical color triplet boson with
fractional electric charge, and non--zero lepton and baryon numbers.
 Such particles are often invoked in extensions of the SM, {\it e.g.}~in
grand unified theories and technicolor models\,\cite{bschrempp}. It is
possible, indeed desirable in some models, that a LQ couple to multiple
lepton and/or quark flavors, thereby providing a mechanism for flavor
violation.

The simplest models involving a flavor--violating leptoquark would
be characterized by three parameters: the leptoquark mass, and
the coupling at each lepton--quark--leptoquark vertex. In order
to avoid models which would involve additional parameters,
we have assumed the following four points:
\begin{enumerate}
\item[1)]
the LQ has ${\mbox{\rm SU}}(3)_C\times{\mbox{\rm SU}}(2)_L\times{\mbox{\rm U}}(1)_Y$
invariant couplings,
\item[2)]
the LQ has either left-- or right--handed couplings, but not both
({\it i.e.}~$\lambda_L\lambda_R=0$),
\item[3)]
the members of each weak--isospin multiplet are degenerate in mass,
\item[4)]
one LQ species dominates the production process.
\end{enumerate}
There are fourteen species of leptoquarks which satisfy these conditions
\cite{buchmuller}.  For fermion number $F\equiv L+3B$ 
($L$ and $B$ denote lepton and baryon number) equal to zero, the
species are denoted \cite{bschrempp}
$S_{1/2}^L$, $S_{1/2}^R$, ${\tilde S}_{1/2}^L$,
$V_0^L$, $V_0^R$, ${\tilde V}_0^R$, and $V_1^L$. 
For $F=2$, they are
$S_0^L$, $S_0^R$, ${\tilde S}_0^R$, $S_1^L$, $V_{1/2}^L$, $V_{1/2}^R$,
 and ${\tilde V}_{1/2}^L$. 
 Here $S$ and $V$ indicate scalar and vector leptoquarks 
respectively, which couple to 
left-- ($L$) or right--handed ($R$) leptons as indicated by the superscript.
The subscript gives the weak isospin of the LQ\footnote{
The tilde differentiates LQ species which differ only in that
one species couples to $u$-type quarks and the other to $d$-type quarks.
See \cite{buchmuller} for details.}.
In $s$-channel reactions, $F=0$ LQ cross sections are higher in 
$e^+ p$ collisions, where they are produced via
$e^+q$ fusion, than in $e^-p$ collisions where $e^-{\overline q}$
fusion occurs. The reverse is true for an $F=2$ leptoquark.

A LQ scenario is defined by the leptoquark species,
by the generations of the quarks which couple to the electron and
to the final--state lepton, and by the final--state lepton flavor.
Hence there are $14\times 3\times3\times2=252$ different LQ scenarios,
each characterized by two dimensionless couplings,
$\lambda_{eq_1}$ and $\lambda_{\lepton q_2}$, defined in figure \ref{fig:stugraph},
which could induce flavor violation. Such LQs
would also mediate flavor--conserving interactions with a final--state
$e$ or $\nu_e$, which are not considered in this paper.

As an illustration, we show in figure \ref{fig:illulimit} 
the present limits \cite{davidson} on $\lambda_{eq_1}\lambda_{\lepton q_2}$
versus LQ mass, $\MLQ$, for reactions which could
proceed through the left--handed scalar isosinglet LQ, $S_0^L$.
Note that each of the limits assumes that specific quark flavors
couple to $e$ and $\mu$. For
example, the most sensitive limit, from $\mu N\to eN$, applies only
for first--generation quarks in both initial and final states.
Also in this figure are the results of direct searches for
leptoquark pair production. Searches in
$e^+e^-$ collisions at LEP\cite{lep} exclude scalar
leptoquarks lighter than about 45 GeV which couple to
$e$, $\mu$, $\tau$, or any neutrino. For leptoquarks
with couplings of electromagnetic strength, masses
below 73 GeV are excluded.
While the LEP experiments did not search for flavor violation,
their non-observation of $e{\overline e}q{\overline q}$,
$\mu{\overline\mu}q{\overline q}$, $\tau{\overline\tau}q{\overline q}$,
or $\nu{\overline\nu}q{\overline q}$
final states kinematically consistent with leptoquark
pair production imply flavor--violating leptoquark mass limits
which are weaker by at most a few GeV.
At the Tevatron \cite{tevatron}, searches in $p{\overline p}$ collisions
have excluded scalar leptoquarks lighter than 131 GeV (96 GeV) for an assumed
branching fraction to $q\mu$ of 100\% (50\%).
These limits on $\MLQ$ are independent of the LQ couplings in most models.

The LQ--induced cross sections for reaction (\ref{epreac}), given in detail in
the appendix, depend on the initial quark density, the couplings, and the
species of LQ involved in the reaction, as well as on the kinematic event
variables $x$ (the Bjorken scaling variable) and
$y$ (the inelasticity). Here $x$ is defined as
$x=-q^2/(2q\cdot P)$ and $y$ as $y=(q\cdot P)/(k\cdot P)$, where $k$, $k'$, and
$P$ are the four--momenta of the initial--state electron, the
final--state lepton and the proton 
respectively, and $q=k-k'$. The square of the 
center--of--mass energies of the electron--proton and
the electron--quark systems are given by $s=(k+P)^2$ and $\shat= xs$
respectively. The remaining Mandelstam variables are given by
$t=-sxy$ and $u=-sx(1-y)$.

For a given coupling, the cross section is largest when the LQ mass,
$\MLQ$, is less than $\sqrt{s}$.  In this case, the LQ is produced
in the $s$--channel, as indicated in figure \ref{fig:stugraph}a.  
Such a leptoquark will appear as a narrow resonance 
in the $x$--distribution peaked at
$x_0\equiv \MLQsq/s$. In the narrow--width approximation
described in the appendix, the cross section for this process
using unpolarized beams can be written
\begin{equation}
  \sigma_{eq_1\rightarrow \lepton q_2}=
  \frac\pi{4s}\,\lambda_{eq_1}^2B_{\lepton q_2}\,q_1(x_0,\MLQsq)\,
  \int dy f(y), \label{lowmasscs}\\
\end{equation}

\[
{\mbox{\rm where}}\; f(y)=\left\{
\begin{array}{lc}
       1 & {\mbox{\rm scalar\, LQ}} \\
6(1-y)^2 & {\mbox{\rm vector\, LQ}},
\end{array}\right.
\]
and $q_1(x,\MLQsq)$ is the quark density in the proton for the
initial--state quark (or antiquark) flavor $q_1$, 
$\lambda_{eq_1}$ is the coupling at the LQ production vertex and
$B_{\lepton q_2}$ is the branching fraction of the LQ to lepton $\lepton$ and quark
flavor $q_2$.  In this process, the final state lepton will have
a transverse momentum (\ptl) of order $\MLQ/2$.

For resonant $s$--channel production, the cross
section for flavor--violating events is proportional to
$\lambda_{eq_1}^2B_{\lepton q_2}$. We will set
limits on this quantity as a function of $\MLQ$.

For the case $\MLQ\gg\sqrt{s}$, either or both $s$-- and $u$--channel
contributions may be important. The corresponding cross sections
can be written as
\begin{eqnarray}
  \sigma_{eq_1\rightarrow lq_2} & = {\displaystyle{s\over32\pi}}
  \left[{\displaystyle{{\lambda_{eq_1}\lambda_{\lepton q_2}}\over{\MLQsq}}}\right]^2 &
  \int dx\,dy\,x q_1(x,\shat) f(y),            \label{scs} \\
  \sigma_{eq_2\rightarrow lq_1} & = {\displaystyle{s\over32\pi}}
  \left[{\displaystyle{{\lambda_{eq_1}\lambda_{\lepton q_2}}\over{\MLQsq}}}\right]^2 &
  \int dx\,dy\,x q_2(x,-u)   f(y),             \label{ucs} 
\end{eqnarray}
\[
{\mbox{\rm where}}\; f(y)=\left\{
\begin{array}{lc}
\frac 12        & s{\mbox{\rm-channel\, scalar\, LQ}}  \\ 
\frac 12(1-y)^2 & u{\mbox{\rm-channel\, scalar\, LQ}}  \\ 
2(1-y)^2        & s{\mbox{\rm-channel\, vector\, LQ}}  \\ 
2               & u{\mbox{\rm-channel\, vector\, LQ}}, \\ 
\end{array}\right.
\]
and the indices $q_1$ and $q_2$ specify the quark flavors
which couple to the electron and the final--state lepton, respectively.  
Here the final--state lepton again will have
large transverse momentum with $\ptl \approx \sqrt{\shat}/2$.

Notice that in the high--mass case, all information about the leptoquark
mass and couplings 
is contained in the quantity ${\lambda_{eq_1}\lambda_{\lepton q_2}}/{\MLQsq}$,
which is the quantity on which
we set limits.  As might be anticipated, other LFV processes 
mediated by leptoquarks, such as
flavor violating meson decays, are sensitive to exactly this quantity.
Hence, our results may be compared directly with prior LFV searches. This
is done in section \ref{results}.

\subsection{$R$--Parity Violating Squarks}\label{RPVsquark}

Squarks ($\tilde{q}$) are the hypothesized supersymmetric partners of
quarks. In supersymmetry (SUSY), $R$--parity is defined as
$R_P=(-1)^{3B+L+2S}$ where $B$, $L$, and $S$ denote baryon and lepton
numbers and spin respectively. This implies 
that $R_P=+1$ for SM particles and $R_P=-1$ for SUSY particles.
If $R$--parity were conserved, SUSY particles would be produced in
pairs and ultimately decay into the lightest supersymmetric particle (LSP),
which would be stable and neutral.
We refer here to this LSP as the photino ($\widetilde{\gamma }$). In
a model with $R$--parity violation, denoted $\rpvio$, single SUSY
particle production would occur and the LSP would decay into SM particles.
Of particular interest for $ep$ collisions are $R$--parity violating
superpotential terms
of the form \cite{Barger} $\lambda'_{ijk}L^i_LQ^j_L{\overline D}^k_R$.
Here $L_L$, $Q_L$, and ${\overline D}_R$ 
denote left--handed lepton and quark doublets and the right
handed $d$-quark singlet chiral
superfields respectively, and the indices $i$, $j$, and $k$ label their
respective generations. Expanded into four-component Dirac notation,
the corresponding terms of the Lagrangian are
\begin{equation}
{\cal L}=\lambda'_{ijk}\left[
   {\tilde\nu}^i_L {\overline d}^k_R d^j_L
 + {\tilde d}^j_L  {\overline d}^k_R \nu^i_L
 + ({\tilde d}^k_R)^* ({\overline\nu}^i_L)^c d^j_L
 - {\tilde e}^i_L {\overline d}^k_R u^j_L
 - {\tilde u}^j_L {\overline d}^k_R e^i_L
 - ({\tilde d}^k_R)^* ({\overline e}^i_L)^c u^j_L \right] + {\mbox{\rm h.~c.}}
\label{eq:Lsquark}
\end{equation}
For $i=1$, the last two terms will
result in ${\tilde u}$ and ${\tilde d}$ production in $ep$
collisions. Identical terms are found in the Lagrangians
for the scalar leptoquarks $\tilde{S}_{1/2}$ and $S_0$,
respectively \cite{Butterworth}.

Lepton--flavor violating $ep$ interactions would occur in a model with
two non-zero couplings $\lambda'_{ijk}$ which involve different
lepton generations.
For example, the process ${\overline e}d\to\tilde{u}^j\to {\overline\mu} d^k$
shown in figure~\ref{susy:graph}a involves the
couplings $\lambda'_{1j1}$ and $\lambda'_{2jk}$. Similarly, non--zero
values for $\lambda'_{11k}$ and $\lambda'_{3jk}$ would lead to the reaction
$eu\to\tilde{d}^k\to\tau u^j$ shown in figure~\ref{susy:graph}b.
Down--type squarks have the additional
decay ${\tilde d}^k\to\nu^id^j$, a mode unavailable to up--type squarks.

The difference between mechanisms involving
$R$--parity violating squarks and leptoquarks is
that the squarks may have additional $R$-parity conserving decay modes
with final--state neutralinos,
such as ${\tilde q}\to q\photino$ (shown in figure~\ref{susy:graph}c)
or with final--state charginos, as in ${\tilde u}\to d {\tilde\chi}^+$.
The branching
ratios $B_{q\photino}$ for the $R_P$--conserving decay $\tilde{q}\to q\photino$
and $B_{ijk}'$ for any $\rpvio$ decay mode are related \cite{Butterworth} by 
\begin{equation}
   {{B_{ijk}'}\over{(\lambda_{ijk}')^2}} =
   {{B_{q\photino}}\over{
     8\pi\alpha}e_{\tilde{q}}^2(1-{m_{\tilde{\gamma}}^2}/{m_{\tilde{q}}^2)^2}},
  \label{eq:translqsq}
\end{equation}
where $\lambda_{ijk}'$ is the $\rpvio$ coupling at the decay vertex,
$\alpha$ is the electromagnetic coupling\footnote{We evaluate
$\alpha$ at the scale $M_Z$ ($\alpha$=1/128)
because $\shat$ is of order $M_Z$ at HERA.},
$e_{\tilde q}$ is the squark charge in units of the electron charge
and the photino and squark masses are $m_{\photino}$ and $m_{\tilde q}$,
respectively. 

Coupling limits for LFV decays of an $S_0^L$ 
leptoquark can be interpreted as ${\tilde d}^k$ 
coupling limits through the correspondence
$\lambda_{eq_1}\sqrt{B_{\lepton q_2}}=\lambda_{11k}'\sqrt{B_{ijk}'}$ 
where $i$ and $j$ are the generations of the LQ decay products $\lepton$ and $q_2$.
Similarly, coupling limits on the $\tilde{S}_{1/2}^L$ LQ can be converted
to limits on couplings to ${\tilde u}^j$ via
$\lambda_{eq_1}\sqrt{B_{\lepton q_2}}=\lambda_{1j1}'\sqrt{B_{ijk}'}$, where 
$i$ and $k$ are the generations of $\lepton$ and $q_2$.

If the stop ($\sTop$) \cite{stop} is lighter than the top quark, then the
$R_P$--conserving decay $\sTop\rightarrow t\photino$
 (figure~\ref{susy:graph}c) will not exist. In the case of $\sTop$,
the correspondence with the coupling limit on $\tilde{S}_{1/2}^L$
is given by
$\lambda_{ed}\sqrt{B_{\lepton q_2}}=\cos\theta_t\lambda_{131}'\sqrt{B_{i3k}'}$
where $\theta_t$ is the mixing angle between the SUSY partners of
the left-- and right--handed top quarks. Over a broad range of possible
stop masses, it is expected that $\cos^2\theta_t\sim0.5$ \cite{stop}.

\section{The ZEUS Detector and Event Simulation}

The main components of the ZEUS detector \cite{STATREP}
used for this analysis
were the uranium--scintillator cal\-or\-i\-me\-ter (CAL)\,\cite{CAL}
and the central tracking detector (CTD) \cite{CTD}.

The CAL, which covers polar angles\footnote{ The ZEUS coordinate system 
is right--handed with the $Z$ axis
pointing in the proton beam direction, hereafter referred to as forward,
and the $X$ axis horizontal, pointing toward the center of HERA. The polar
angle $\theta$ is taken with respect to the proton beam direction from the
interaction point.} between $2.2^{\circ }$ and $176.5^{\circ}$,
is divided into forward (FCAL), barrel (BCAL), and rear (RCAL)
parts. Each part is further subdivided into towers which are longitudinally
segmented into electromagnetic (EMC) and hadronic (HAC)
sections. In depth, the EMC is one interaction length; the HAC sections 
vary from six to three interaction lengths, depending on polar angle.
Under test beam
conditions \cite{CAL}, the calorimeter has an energy resolution of
$\sigma_E(\mbox{\gev})=0.18\,\sqrt{E(\mbox{\gev})}$ for electrons and
$\sigma_E(\mbox{\gev})=0.35\,\sqrt{E(\mbox{\gev})}$ for hadrons.  In this analysis, 
only cells with energies above  noise suppression thresholds
(60 MeV for EMC, 110 MeV for HAC) were used.

A superconducting coil located inside the CAL provides 
a 1.43 Tesla magnetic field parallel to the beam axis in which 
the charged particle tracking system operated.
The interaction vertex is reconstructed with a resolution of 4 mm
(1 mm) along (transverse to) the beam direction. 
The muon detection system\,\cite{zeusmuon} was used to check the
efficiencies and the background estimates for the primary
muon identification, which used only the CAL and the CTD.
The muon detectors are also divided into three sections covering the
forward, barrel, and rear regions. In the barrel and rear sections,
which were used for this analysis,
the detectors consist of eight layers of limited streamer tubes,
four layers on each side of the 80 cm thick magnetized iron yoke.
Luminosity was measured \cite{LUMI} from the rate
of brems\-strah\-lung events ($ep\rightarrow ep\gamma $) detected by
a photon calorimeter (LUMI) located downstream of the main
detector. The luminosity is known to 3\% for the $e^-p$ data
and to 2\% for the $e^+p$ data.

To evaluate detection efficiencies, we have simulated flavor--violating
LQ processes
using a modified\footnote{The final--state electron and quark from these
generators were replaced by the appropriate lepton ($\mu$ or $\tau$),
and quark species, before the simulation of parton showering and
fragmentation. Both $s$-- and $u$--channel exchange
contributions were included.} version of {\sc pythia} \cite{pythia} and
also with {\sc lqmgen} which is based on the
differential LQ cross-sections given in
\cite{buchmuller}. The calculations included initial state brems\-strah\-lung.

For background estimation, charged--current (CC) and neutral--current (NC)
deep--inelastic
scattering (DIS) events with electroweak radiative corrections were 
simulated using {\sc lepto} \cite{lepto} interfaced to {\sc heracles}
\cite{heracles} via
{\sc django} \cite{django}.  The MRSA \cite{mrsa} parton density
parameterization was used.  The hadronic final state was simulated using 
{\sc ariadne} \cite{ariadne} and {\sc jetset} \cite{pythia}.

Photoproduction processes were simulated using {\sc herwig} \cite{herwig},
and photoproduction of $c\overline{c}$ and $b\overline{b}$ pairs by
{\sc pythia} and {\sc aroma} \cite{aroma}.  
The processes $\gamma\gamma\to\mu^+\mu^-$ and
$\gamma\gamma\to\tau^+\tau^-$ were generated using {\sc zlpair}
\cite{zlpair}. Finally, production of $W$ bosons was simulated using {\sc
epvec} \cite{epvec}.

All generated events were passed through a {\sc geant} \cite{geant} based
detector simulation which tracked final state particles 
and their decay and interaction products through the entire detector.
The simulated events were processed with the same
analysis programs as the data.

\section{Trigger and Analysis}

The signature of LFV events ($e+p\rightarrow \lepton+X$) in this experiment
is an isolated $\mu $ or $\tau$
of high transverse momentum, $\ptl\sim \sqrt{\shat}/2$, balanced
by a jet of hadrons.

\subsection{Search Strategy}

Our search strategy relies on the fact that the LFV signal events will
almost always have a large net transverse momentum $\ptmiss$
measured in the calorimeter. We reconstruct $\ptmiss$ 
as $\ptmiss =(P_X^2+P_Y^2)^{1/2}$.
Here $P_X=\sum_i E_i \sin(\theta_i) \cos(\phi_i)$ and 
$P_Y=\sum_i E_i \sin(\theta_i) \sin(\phi_i)$ where the sums run over all
calorimeter cells and $E_i$, $\theta_i$, and $\phi_i$ are the energy,
polar angle and azimuthal angle of cell $i$, calculated using the
reconstructed event vertex. We also reconstruct
the azimuth of the missing transverse momentum, determined from
$\cos\phi\miss=-P_X/\ptmiss$ and $\sin\phi\miss=-P_Y/\ptmiss$.

A high energy muon is a  minimum ionizing particle, 
typically producing a measured energy of about 2 GeV in the calorimeter.
If the much larger muon transverse momentum, $P_t^\mu$, is balanced by a jet
of hadrons, then $\ptmiss\approx P_t^\mu$.
Thus the signature for such an event would be a large $\ptmiss$
and a high momentum track which points to an isolated calorimeter
cluster with approximately 2 GeV of energy at an azimuthal angle $\phi\miss$.

A final--state $\tau$ decays promptly to a small number of
charged particles (1 or 3, 99.9\% of the time), zero or more
neutral hadrons, and at least one neutrino. Since the $\tau$ mass is small
($m_\tau =1.78$ GeV) compared to its transverse momentum,
the $\tau$ decay products will be collimated in a cone of
opening angle $\sim 0.03$ radians.
For events in which the $\tau$ decays via
$\tau\rightarrow\mu\nu{\overline\nu}$,
the experimental signature will be similar
to an event with a final state muon except with $P_t^\mu<\ptmiss$.
If the $\tau$ decays via $\tau\rightarrow e\nu{\overline\nu}$,
the event will be characterized by large $\ptmiss$ (due to the 
undetected neutrinos) and the presence of 
a high--transverse--momentum electron with azimuth $\phi\miss$.
Finally, in the case of a hadronic $\tau$ decay, we would again
see a large $\ptmiss$ due to the neutrino, and a compact
hadronic cluster with 1 or 3 tracks, also at azimuth $\phi\miss$.

\subsection{Trigger}
Data were collected with a three--level trigger system \cite{STATREP}.  
Since the signature which we are seeking is one with missing transverse
momentum measured in the calorimeter, our triggering scheme was
largely calorimeter based.
The first--level triggers used
net transverse energy,
missing transverse energy, as well as EMC energy sums
in the calorimeter. The thresholds were well below the offline requirements.
The second--level trigger rejected backgrounds
(mostly $p$--gas interactions and cosmic rays) for which the calorimeter timing
was inconsistent with an $ep$ interaction. Events were accepted
if $\ptmiss$ exceeded 9 GeV and either a track was found in the
CTD or at least 10 GeV was deposited in the FCAL. The
latter alternative was intended to accept events with jets which are
too forward for the tracks to be observed in the CTD.
The third--level trigger
applied stricter timing cuts and also pattern recognition algorithms to reject
cosmic rays.

\subsection{Leptoquark Mass Reconstruction}\label{LQMassRec}
For $\MLQ< 300$ GeV, the leptoquark is produced
as an $s$-channel resonance and consequently, the
invariant mass distribution of the $q\lepton$ final state
is sharply peaked at $\MLQ$.
When searching for a leptoquark of a given mass,
the expected background can be reduced by requiring the
reconstructed $q\lepton$ mass to be consistent with $\MLQ$.

We reconstruct the leptoquark mass as follows
using a simple ansatz based on three approximations:
1) the four--momentum of all final state muons and neutrinos can be
represented by a single massless pseudoparticle;
2) the contribution of the proton remnant to the reconstructed mass can be
ignored; and 
3) no energy escapes through the rear beam hole.

The four--momentum of  the invisible pseudoparticle, $P\invis$,
is related to the net four--momentum $P=(E,P_X,P_Y,P_Z)$ measured in the
calorimeter as $P_X\invis=-P_X$, $P_Y\invis=-P_Y$, and
$E-P_Z+E\invis-P_Z\invis=2E_e$ where $E_e$ is the electron beam energy.
The reconstructed leptoquark mass is given by
 $(M_{\scriptscriptstyle LQ}^{\scriptscriptstyle rec})^2=(P+P\invis)^2$.

We have applied this mass reconstruction to simulated LFV events and
determined two functions, $\MLQr(\MLQ)$ and $\sigLQr(\MLQ)$ which
give the mean and the standard deviation of a Gaussian fit to
the reconstructed mass distribution as a function of the
true $\MLQ$. Studies of simulated LQ signals indicate that
the mass resolution improves from about 13\% at
$\MLQ=100$ GeV, to about 6\% at $\MLQ=250$ GeV.

\subsection{Event Selection}
The most important offline selection requires that
$\ptmiss$ exceed 12 GeV.
The initial event selection is designed to accept all
$ep$ collisions which meet this condition, while efficiently
rejecting the high--rate
backgrounds from cosmic rays, proton--gas interactions, off--beam protons,
and beam--halo muons. Triggers from these backgrounds usually do not have
a reconstructed vertex. In cases where a spurious vertex is reconstructed,
it typically is made from a small number of
low--momentum spiraling tracks which do not intersect with the beam line. 
Unlike $ep$ collisions, for which the  distribution of $Z$ vertex position
is centered at $Z=0$ with an r.~m.~s. width of 12 cm, the spurious vertices
have a $Z$ distribution which is roughly uniform.
In cases of protons colliding with residual
gas in the beam pipe, or with the beam--pipe itself, the low--multiplicity
spurious vertex is  accompanied by a large number (10 to 100) of
tracks which are not correlated with the vertex.
Occasionally a cosmic ray or a beam--halo
muon will coincide with an $ep$ interaction which provides
the reconstructed vertex. In these cases the vertex tracks are typically
of quite low momentum (${\cal O}(100$ MeV)).

    In order to remove such backgrounds, we require that
a vertex is reconstructed and that it lie within 50 cm
of the nominal interaction point.
We define $N_{trk}$ to be the total number of
reconstructed tracks, $N_{good}$ to be the number of tracks with
transverse momentum $P_t>300$ MeV and
a distance of closest approach to the beam--line
of less than 1.5 cm,  and $N_{vtx}$ to be the number of tracks forming
the vertex. 
We require $N_{good}\ge 1$ and $N_{good}\ge 0.05N_{trk}$.
In order to reject proton--induced background, for which the energy
deposited in the calorimeter is concentrated at small polar angles,
we remove events with $P_Z/E>0.8 (0.94)$\footnote{
Here $P_Z$ and $E$ are reconstructed from the calorimeter cells in a
manner similar to the components $P_X$ and $P_Y$ described above.}
if $N_{trk}-N_{vtx}\ge 80$ (20).
In addition, we require the timing of each calorimeter cluster
with energy above 2 GeV to be consistent with an $ep$ interaction.
To reduce the cosmic ray background, we apply an algorithm which rejects
events in which the pattern of calorimeter energy deposits is consistent with
a single penetrating particle traversing the detector.

The 175 events which passed these cuts were visually examined and
29 events clearly initiated by cosmic rays, muons in the beam halo, or
anomalous photomultiplier discharges were removed,
leaving 146 $ep$ collision events.
These events were divided into two classes:
those for which no isolated electron with energy
$E_e>10$ GeV was found in the calorimeter (class $\noe$); and those for
which such an electron was found (class $e$).
The following selection cuts, which were developed in Monte Carlo
studies, were applied to each sample in order to eliminate SM backgrounds.
\begin{description}
\item[$\noe$:]\hspace{6mm} There were 114 events with no isolated electron.
      Six events were rejected
      because an electron of more than 5 GeV was observed in the luminosity
      electron calorimeter and they were thus recognized to be
      background due to photon--proton ($\gamma p$) collisions.
      This left 108 $\noe$ events, which agrees well
      with the Monte Carlo estimates of 100 CC DIS
      events and 15 $\gamma p$ and $\gamma\gamma$ events.
      The $\ptmiss$ distribution of this $\noe$ sample is compared with the
      Monte Carlo prediction in figure \ref{fig:ptclass12}a.
      The $\noe$ sample serves as the source of flavor violation candidates
      with a muon in the final state, as well as of candidates with
      final--state
      $\tau$'s which subsequently decay via 
      $\tau\rightarrow \mu \overline\nu  \nu$ or
      $\tau\rightarrow \nu +$hadrons. 
\item[$e$:]\hspace{6mm} There were 32 events which contained an
      isolated electron. In order
      to reject NC DIS background, for which the electrons
      are concentrated at large polar angles, we required the electron
      polar angle to be less than 100$^\circ$. After this cut, 12 events
      remained in the $e$ sample, compatible with the Monte Carlo prediction
      of 14 NC DIS events. 
      The $\ptmiss$ distribution of these remaining events 
      is compared with the
      Monte Carlo prediction in figure \ref{fig:ptclass12}b.
      LFV candidates with a final--state $\tau$
      which decays via $\tau\rightarrow e\overline\nu\nu$
      (18\% branching fraction) were sought in this sample.
\end{description}

The final cuts rely on a clustering algorithm\footnote{
The clustering algorithm joins each cell with its highest energy
neighbor, thus producing one cluster for each cell which has more
energy than any of its neighbors. Two cells are defined as neighbors
if they are in towers which share a face or an edge.
Cells on the forward or rear edges 
of the BCAL are also neighbors with the FCAL or RCAL cells which are
behind them, as viewed from the interaction point.} 
which assigns each calorimeter cell above noise threshold to one and only one 
cluster. Each cluster is
characterized by its energy, $E\clu$, as well as the energy--weighted
mean azimuth, $\phiclu$, and 
pseudorapidity, $\eta\clu=-\ln[\tan(\theta\clu/2)]$. 
We expect the final--state lepton in a LFV event to 
produce a single isolated cluster. To decide if a cluster is
isolated, we examine the set of all calorimeter cells which are
within 0.8 units in $\eta\phi$ of the cluster
($[(\phiclu-\phi_{cell})^2+(\eta\clu-\eta_{cell})^2]^{1/2}<0.8$).
A cluster is defined to be
{\em isolated} if the summed energy of all calorimeter cells in
this set which do not belong to the cluster is below 2 GeV.
For each cluster, we also compute $\phiclubar$, which is
defined as the energy weighted mean azimuth of all cells 
in the entire calorimeter, except for those assigned to the cluster.
Note that for LFV events $\phiclubar$ differs only slightly from
$\phi\miss$ which is computed using all calorimeter cells.
A cluster is said to be {\em \ptmiss--aligned} if it 
satisfies the inequality: $\cos(\phiclu-\phiclubar)<\cos170^\circ$. This
ensures that the isolated cluster is opposite in azimuth to the rest of the
energy in the calorimeter.

To enter the final sample for $\mu q$ or
$\tau q$ final states, an event must
satisfy the criteria of one of four selections, described below.
\begin{description}
\item[$\mu$ {\mbox{\rm or}} $\tau\rightarrow\mu$:] In a class $\noe$ event,
      there must exist an isolated \ptmiss--aligned cluster
      with energy 0.5 GeV $<E\clu<6$ GeV and at most 80\% of its
      energy in the electromagnetic layer of the calorimeter.
      It must have exactly one matching track\footnote{
A matching track is defined such that the distance of closest
approach between the extrapolated track and the calorimeter
cluster is less than 30 cm.} 
      and that track must have momentum exceeding 20 GeV.
      The efficiency\footnote{All quoted efficiencies include the
      trigger efficiency.} to satisfy these cuts for scalar
      (vector) leptoquarks which decay to $\mu q$ decreases with LQ mass from
      74\% (78\%) at $\MLQ=80$ GeV to 31\% (50\%) at $\MLQ=260$ GeV.
      The background estimate for the $\mu$ selection was 0.1 events from 
      the inelastic process $\gamma\gamma\rightarrow\mu^+\mu^-$.
      Zero events were observed in the data.

      To check the efficiency of the muon selection,
      we performed an independent event selection which did not use the CAL,
      but required a track in the barrel or rear
      muon chambers which was matched to a CTD track with a
      transverse momentum of at least 5 GeV. In order to select
      events with isolated muons, we rejected events which had an electron
      found in the calorimeter or had more than three tracks fitted to the
      vertex. A total of 15 events, which contained 17 CTD--matched muon
      chamber tracks passed these cuts. This number agrees with the Monte Carlo
      estimate of 20 events from $\gamma\gamma\rightarrow\mu^+\mu^-$.
      All 17 tracks were matched to an isolated
      calorimeter cluster which passed the cuts described above
      (except for the $\ptmiss$-alignment).
      
\item[$\tau\rightarrow e$:] In a class $e$ event, the isolated electron must
      be \ptmiss--aligned. The efficiency to satisfy these cuts for scalar
      (vector) leptoquarks with final--state $\tau q$, and the subsequent decay
      $\tau\rightarrow e\overline\nu\nu$, rises with LQ mass from
      23\% (17\%) at $\MLQ=80$ GeV to 75\% (75\%) at $\MLQ=260$ GeV.
      The background estimate for the $\tau\rightarrow e$ selection
      was 0.2 events from NC DIS.
      Zero events were observed in the data.

\item [$\tau\rightarrow$ {\mbox{\rm hadrons:}}] In a class $\noe$ event, there
      must exist an isolated \ptmiss--aligned cluster with
      $E\clu>10$ GeV, which has either 1 or 3 matching tracks.
      At least one track must have a momentum exceeding 5 GeV.
      The efficiency to satisfy these cuts for scalar
      (vector) leptoquarks with final--state $\tau q$ and hadronic
      $\tau$ decay rises with leptoquark mass from
      15\% (12\%) at $\MLQ=80$ GeV to 39\% (47\%) at $\MLQ=260$ GeV.
      We estimated a background of 0.4 events for this
      selection coming from CC DIS (0.2 event),
      $\gamma\gamma\rightarrow\mu^+\mu^-$ (0.1 event), and
      $c{\overline c}$ production (0.1 event). We observed
      zero events.

\item [$\ptmiss>80$ {\mbox{\rm GeV:}}] Leptoquarks with mass in the range
      200 GeV$<\MLQ<300$ GeV
      would be  strongly boosted in the forward direction so that
      the final state $\mu$ or $\tau $ would often have polar angle
      less than 10$^\circ$.  In such cases, the final--state lepton
      would be outside the CTD acceptance and would consequently fail
      the track matching cuts. In order to maintain high efficiency
      at these masses, we accept {\em any} event from either class $e$ or
      class $\noe$ for which $\ptmiss>80$ GeV. 
      For 240 GeV leptoquarks which decay to $\mu q$,
      accepting events with $\ptmiss>80$ GeV increases the overall
      acceptance from 36\% to 69\% for scalars, and from 53\% to 76\%
      for vectors.
      For the $\ptmiss>80$ GeV selection, we estimated a background of
      1.0 events from CC DIS and we observed zero events.
\end{description}

For the low--mass leptoquark search ($\MLQ<300$ GeV),
one additional cut was applied, which,
in contrast to all cuts described above,
depends on $\MLQ$, the mass of the LQ being searched for.
The leptoquark mass was reconstructed using the method described
in \ref{LQMassRec} and we required that the reconstructed
mass must lie within $3\sigLQr(\MLQ)$ of $\MLQr(\MLQ)$.

\section{Results}\label{results}

With no candidate events for LFV found in either the $e^-p$
or $e^+p$ data samples with integrated luminosities $\lum_{e^-}=0.84$ \ipb
and $\lum_{e^+}=2.94$ \ipb, we set upper limits on
the couplings of the various LFV processes described in section \ref{LFVsce}.
The upper limit on the coupling $\lambda$ is obtained
from the relation $N=\sum_{i=e^+,e^-} \lum_i\epsilon_i\sigma_i(\lambda)$,
where $\epsilon$ is the
efficiency, $\sigma(\lambda)$ is the cross section for a coupling $\lambda$,
and $N$ is the Poisson 95\% confidence level (CL) upper limit \cite{databook}
on the number of events. The signal efficiencies and background
estimates were determined by Monte Carlo studies. We estimate the
systematic uncertainties in the efficiencies to be 5\%. Cross
sections were calculated using the formulae given in the appendix
and the GRV--HO \cite{grv} parton density parameterization.
Cross sections calculated using the MRSH \cite{mrsh} parameterization
differ in magnitude by less than 12\% for $u$, $d$, $s$, or $c$ quarks
in the initial state, and by less than 19\% for initial--state $b$ quarks.

\subsection{Low-Mass Leptoquark Limits (\mbox{$M_{LQ}< 300$ GeV})}
\begin{figure}
  \centerline{\psfig{figure=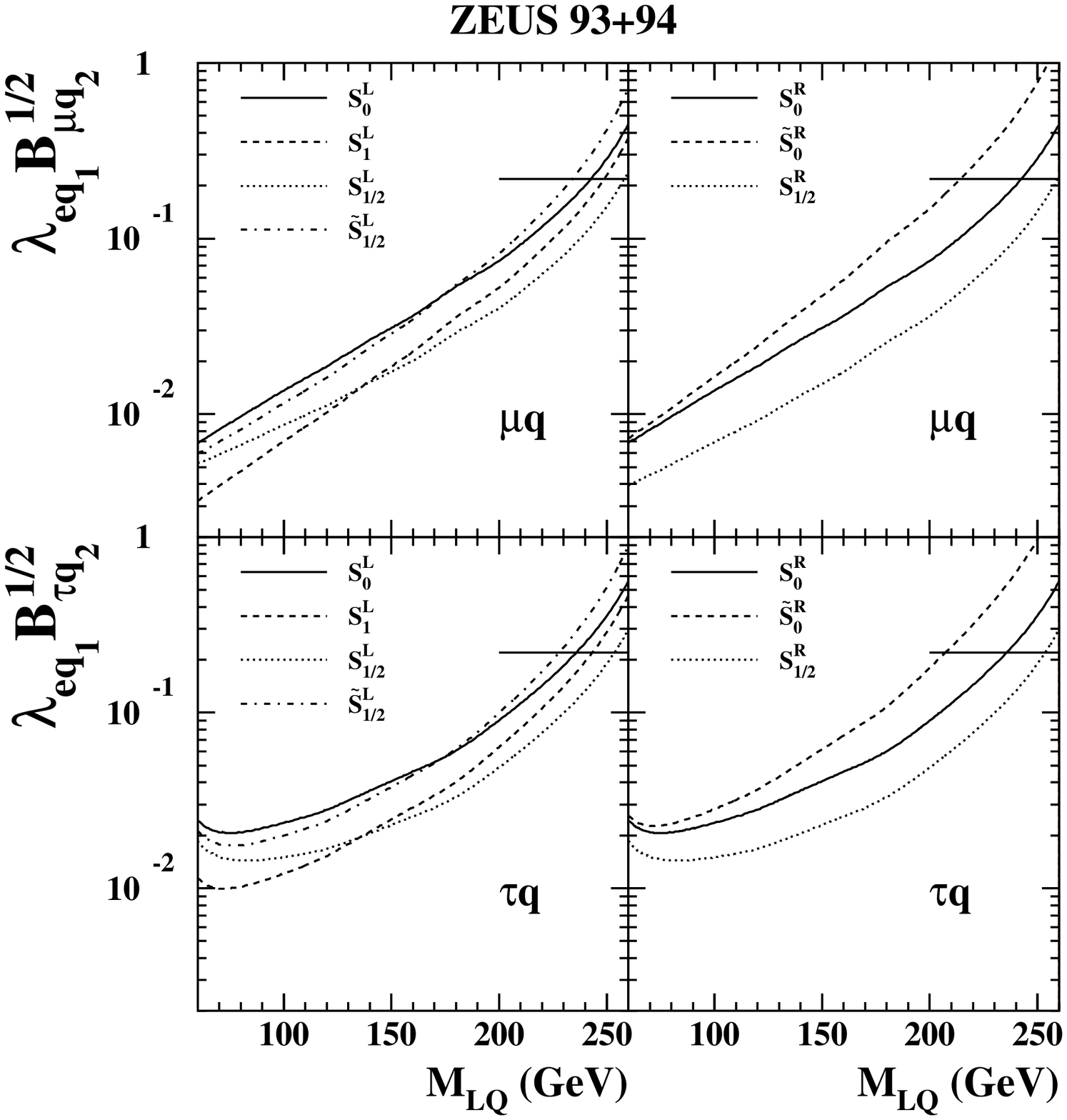,width=20cm,height=20cm}}
  \caption{
The upper limit on the coupling at the production vertex ($\lambda_{eq_1}$)
times the square root of the branching
fraction to the $\mu q$ or $\tau q$ final state ($B_{\lepton q_2}$)
{\it vs.}~leptoquark mass $M_{LQ}$, at $95\%$ CL for scalar leptoquarks.
The horizontal line indicates nominal electromagnetic coupling
($\lambda_{eq_1}^2=4\pi\alpha=4\pi/128$) and $B_{\lepton q_2}=0.5$. }
  \label{fig:scalar}
\end{figure}

\begin{figure}
  \centerline{\psfig{figure=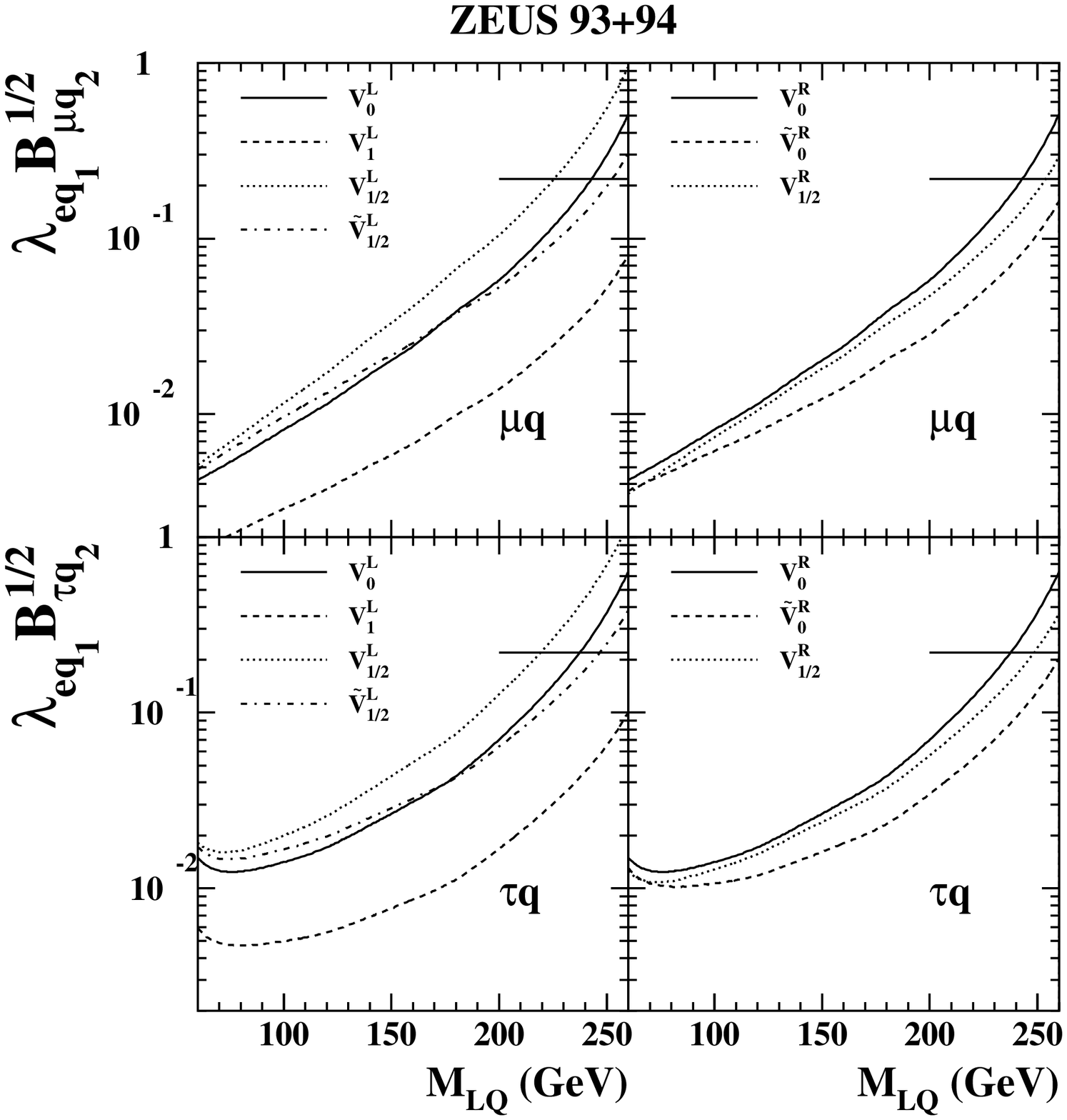,width=20cm,height=20cm}}
  \caption{
The upper limit on the coupling at the production vertex ($\lambda_{eq_1}$)
times the square root of the branching
fraction to the $\mu q$ or $\tau q$ final state ($B_{\lepton q_2}$)
{\it vs.}~leptoquark mass $M_{LQ}$, at $95\%$ CL for vector leptoquarks.
The horizontal line indicates nominal electromagnetic coupling
($\lambda_{eq_1}^2=4\pi\alpha=4\pi/128$) and $B_{\lepton q_2}=0.5$. }
  \label{fig:vector}
\end{figure}

In the case of low--mass leptoquarks, we calculate
upper limits on $\lambda_{eq_1}^2B_{\lepton q_2}$ using equation \ref{lowmasscs}.
Since our limits are largely independent of the
final state quark type (as long as the top quark is not involved), we show
in figures~\ref{fig:scalar} and \ref{fig:vector} the upper limits on
$\lambda _{eq_1}\sqrt{B_{\lepton q_2}}$, for scalar and vector leptoquarks where
$q_1$ is a first--generation quark and $q_2$ is the final--state
quark of any generation (except top).
 The limits for $\lepton=\mu$ and for $\lepton=\tau$ are shown separately
as a function of LQ mass for the various scalar
and vector LQ species. We note that for several LQ species,
we probe coupling strengths as small as
$\lambda _{eq_1}^2/4\pi\approx 10^{-3}\alpha$ for 
$\MLQ=100$ GeV and $B_{\lepton q_2}=0.5$.

\begin{figure}
  \centerline{\psfig{figure=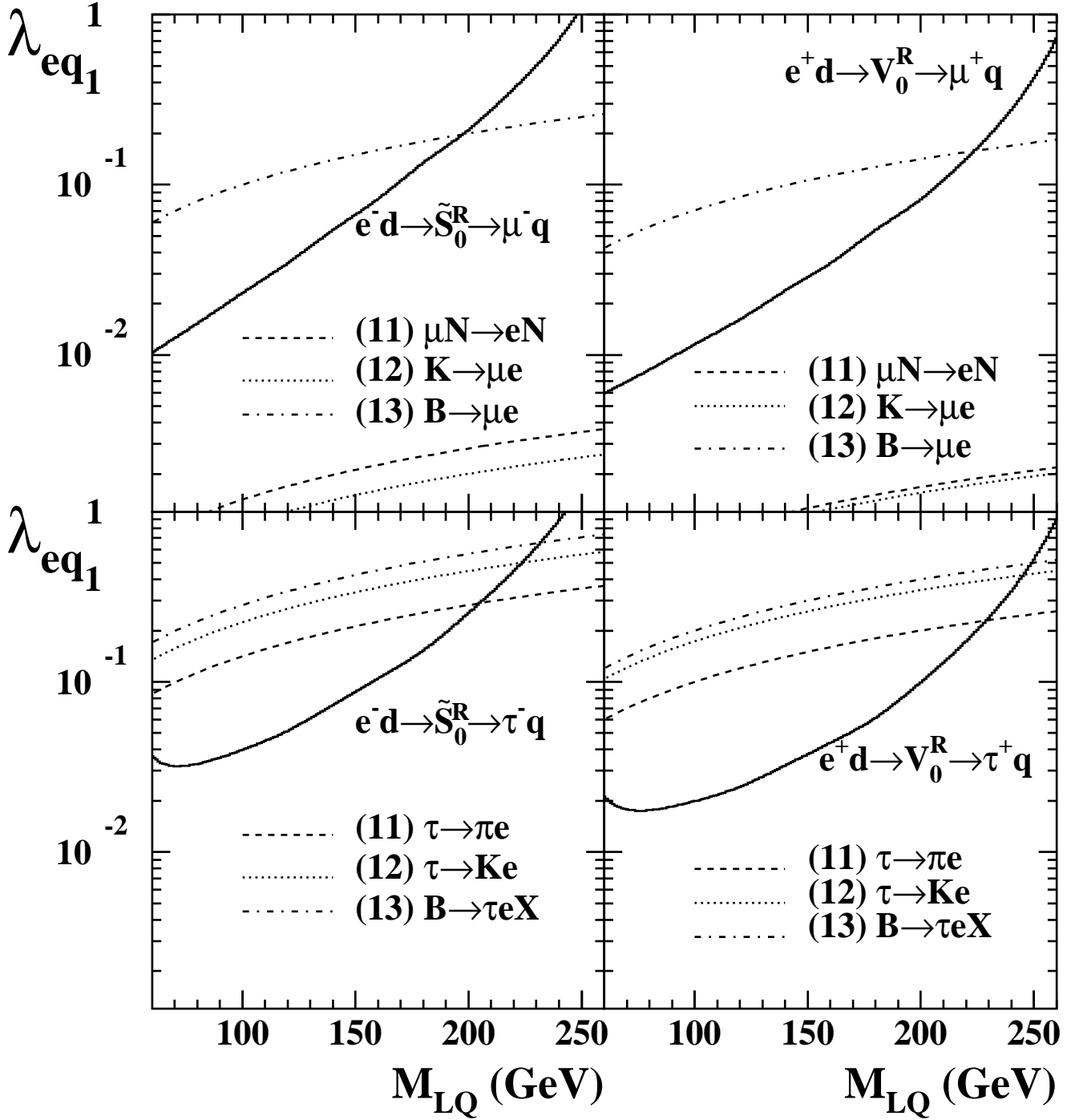,width=20cm,height=20cm}}
  \caption{
The $95\%$ CL upper limits on $\lambda_{eq_1}$ {\it vs.}~leptoquark
mass $M_{LQ}$, for selected LQ species which decay to $\lepton q$, 
where $\lepton=\mu$ (above) or $\tau$ (below),
assuming $B_{\lepton q_2}=0.5$.  The solid curves are ZEUS results and
the various broken curves show existing limits \protect\cite{davidson}.
Paired numbers in parentheses indicate the generations of the quarks
which couple to $e$ and $\lepton$ respectively.}
  \label{fig:lqlimit}
\end{figure}

In figure~\ref{fig:lqlimit} we compare these limits on LFV with those from
previous searches for two representative LQ species,
$\tilde{S}_0^R$ and $V_0^R$. Assuming that $B_{\lepton q_2}=0.5$,
we plot as a solid curve the upper limit on $\lambda_{eq_1}$ as a
function of the LQ mass. Curves are shown for both $\mu$ (upper plots)
and $\tau$ (lower plots) final states. 
In contrast with many other limits on LFV, the coupling limits from
this experiment apply to final--state quarks
of any generation (except top). The various broken
curves are  low--energy limits quoted from reference~\cite{davidson}.
For each of these curves, the pairs of numbers in parentheses
denote the generations of quarks which couple to $e$ and $\lepton$.
Coupling limits for $B_{\lepton q}\neq 0.5$ can
be obtained by multiplying the limit on $\lambda_{eq_1}$ plotted
in figure~\ref{fig:lqlimit} by
$\sqrt{0.5/B_{\lepton q}}$. We emphasize two important implications of
figure~\ref{fig:lqlimit}: 

\begin{enumerate}
\item  The ZEUS limits on $ed\rightarrow \mu b$ via $\tilde{S}_0^R$
($V_0^R$) for $\MLQ<\sqrt{s}$ supersede previous upper bounds \cite{btomue} from
$B\rightarrow \mu \overline{e}$, for $\MLQ$ below 200 GeV (220 GeV).
On the other hand, the limits from $\mu$ conversion in titanium\,\cite{Tiexp}
and from forbidden $K$
decays\,\cite{rareK} which involve only first and second generation quarks
are much stronger than the corresponding ZEUS limits.

\item  For $\MLQ$ below 200 GeV, the ZEUS limits on
$ed\rightarrow \tau q_2$ through $\tilde{S}_0^R$
and $V_0^R$ supersede previous limits from
$\tau \rightarrow \pi e$ \cite{tautopie},
$\tau \rightarrow Ke$ \cite{tautoKe}, and
$B    \rightarrow \tau eX$ \cite{btomue},
for $q_2=d,s,$ and $b$, respectively.
\end{enumerate}

Figure~\ref{fig:lqlimit} illustrates examples in which the existing
low--energy limits, though less stringent than the ZEUS limits at low
$\MLQ$, become more stringent at higher masses.
As described in the next section, this is not always the case.

\begin{figure}
  \centerline{\psfig{figure=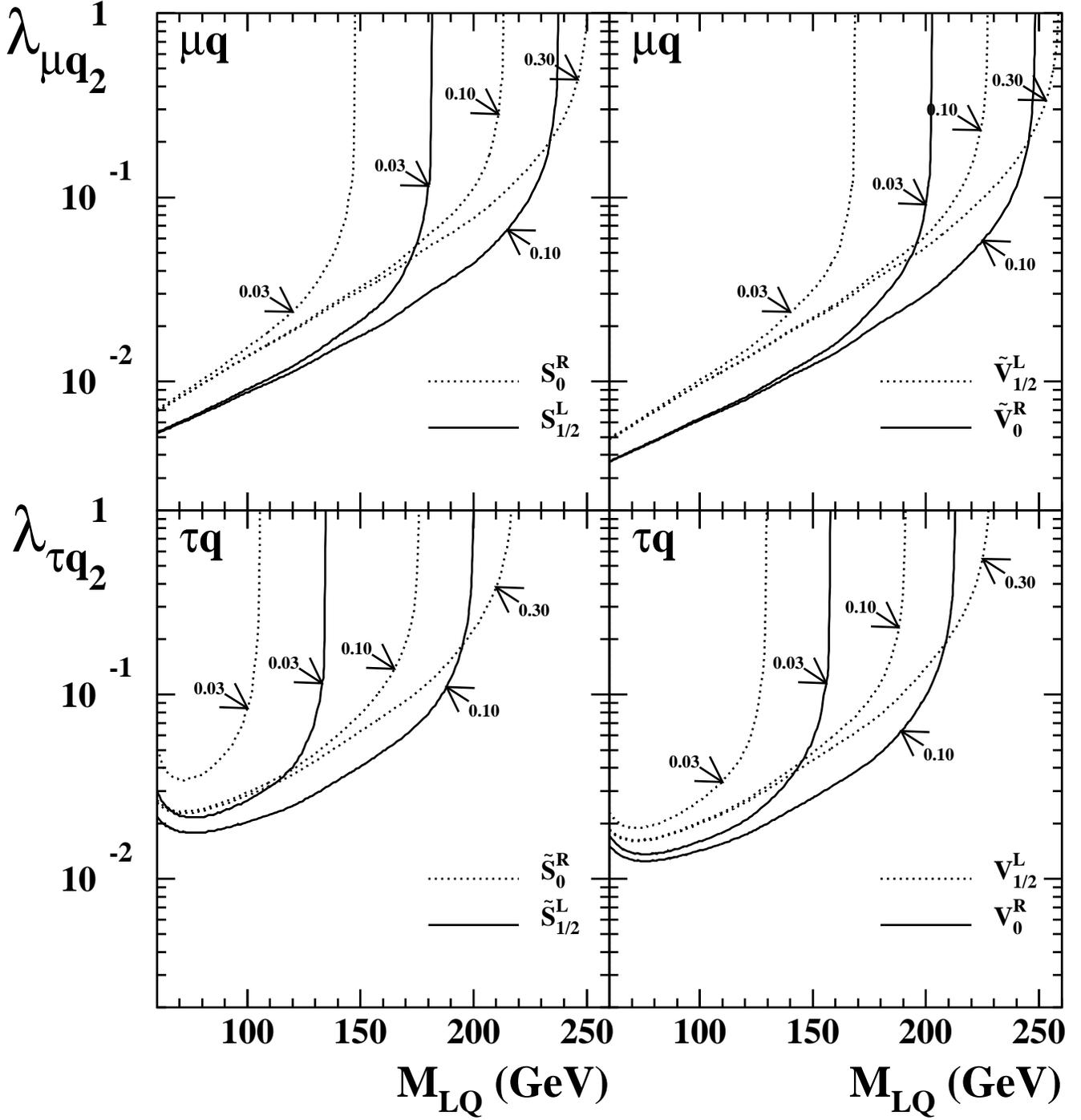,width=20cm,height=20cm}}
  \caption{
The upper limit on the coupling at the decay vertex ($\lambda_{\lepton q_2}$)
{\it vs.}~leptoquark mass $M_{LQ}$, 
for several values of the first--generation coupling at the production
vertex ($\lambda_{eq_1}$). Each curve is labeled by the value of
$\lambda_{eq_1}$.
The dotted curves are for $F=2$ leptoquarks and the
solid curves are for $F=0$ leptoquarks.}
  \label{fig:h1plot}
\end{figure}

An alternative approach to setting limits, which was employed
in reference \cite{H1} is to assume that the branching ratio
$B_{\lepton q_2}$ is given\footnote{This formula for
$B_{\lepton q_2}$ assumes that the leptoquark does not couple to neutrinos.}
by $\lambda_{\lepton q_2}^2/(\lambda_{eq_1}^2+\lambda_{\lepton q_2}^2)$,
and to set limits
on $\lambda_{\lepton q_2}$ for a fixed value of $\lambda_{eq_1}$.
Such
limits are shown in figure \ref{fig:h1plot}. For $F=0$ LQs,
our limits are similar to those of reference \cite{H1}, while
for $F=2$ LQs, the ZEUS limits are stronger due to
inclusion of $e^-p$ data.

Finally, a third way to illustrate the sensitivity is to assume that the LQ
couplings have electromagnetic strength
($\lambda_{eq_1}^2 =\lambda_{\mu q_2}^2 =4\pi\alpha$
for LQs which couple to $e$ and $\mu$)
and to determine a lower limit on the allowed LQ mass.
Such limits are shown in table~\ref{tab:mlimsum}. 
For scalar leptoquarks, lower mass limits between 207 GeV and 259 GeV are set.
Somewhat stronger mass limits, between 219 GeV and 272 GeV, are set
on vector leptoquarks for which both the production cross section and
detection efficiency are higher.

\subsection{High--Mass Leptoquark Limits (\mbox{$M_{LQ}\gg 300$ GeV})}

For high--mass leptoquarks,
the cross section is proportional to the square of
$\Psi_{eq_1\lepton q_2}\equiv\lambda _{eq_1}\lambda _{\lepton q_2}/{\MLQsq}$,
and a factor which does not depend on either the leptoquark couplings or mass
(see equations~\ref{scs} and \ref{ucs}).
This is also true of rates for lower energy forbidden
processes\,\cite{davidson}. 
For a given limit on $\Psi_{eq_1\lepton q_2}$, the limit on the
product $\lambda _{eq_1}\lambda _{\lepton q_2}$
is proportional to $\MLQsq$. As $\MLQ$ increases, the upper limit
on the product of the couplings eventually exceeds unity and the perturbation
expansion, on which the formulae in the appendix are based, breaks down.
Even so,
the parameter $\Psi_{eq_1\lepton q_2}$ serves as a reasonable figure of merit
for experimental comparisons.

Tables \ref{tab:muqF2}, \ref{tab:muqF0}, \ref{tab:tauqF2} and \ref{tab:tauqF0} 
summarize the 95\% CL upper bounds on $\Psi_{eq_1 \lepton q_2}$,
in units of $10^{-4}$ GeV$^{-2}$ from this experiment and also from previous
experiments \cite{davidson}.
Here $q_1$ and $q_2$ are the generation indices of the
quarks which couple to $e$ and to $\lepton$ respectively\footnote{
Certain entries in these tables have been corrected and/or updated from
reference~\cite{davidson} after consultation with the authors~\cite{privcom}.}.
Two important characteristics of these tables are summarized below.

\begin{enumerate}
\item  In the $e\leftrightarrow \mu $ case, for LQ species
$V_{1/2}^L$, $\widetilde{V}_{1/2}^L$, $S_{1/2}^L$, or $\widetilde{S}_{1/2}^L$,
the limits from this experiment supersede prior limits in some cases
where heavy quark flavors are involved,

\item  For the $e\leftrightarrow \tau$ case, we also improve upon
existing limits for the same LQ species as in point 1.
In addition, because the existing limits on
$e\leftrightarrow\tau$ are much weaker than those for $e\leftrightarrow\mu$,
the ZEUS limits are the most stringent for several additional LQs which
couple to $c$ or $b$ quarks.
\end{enumerate}

\subsection{Limits for $\rpvio$ Squarks}

\begin{figure}
  \centerline{\psfig{figure=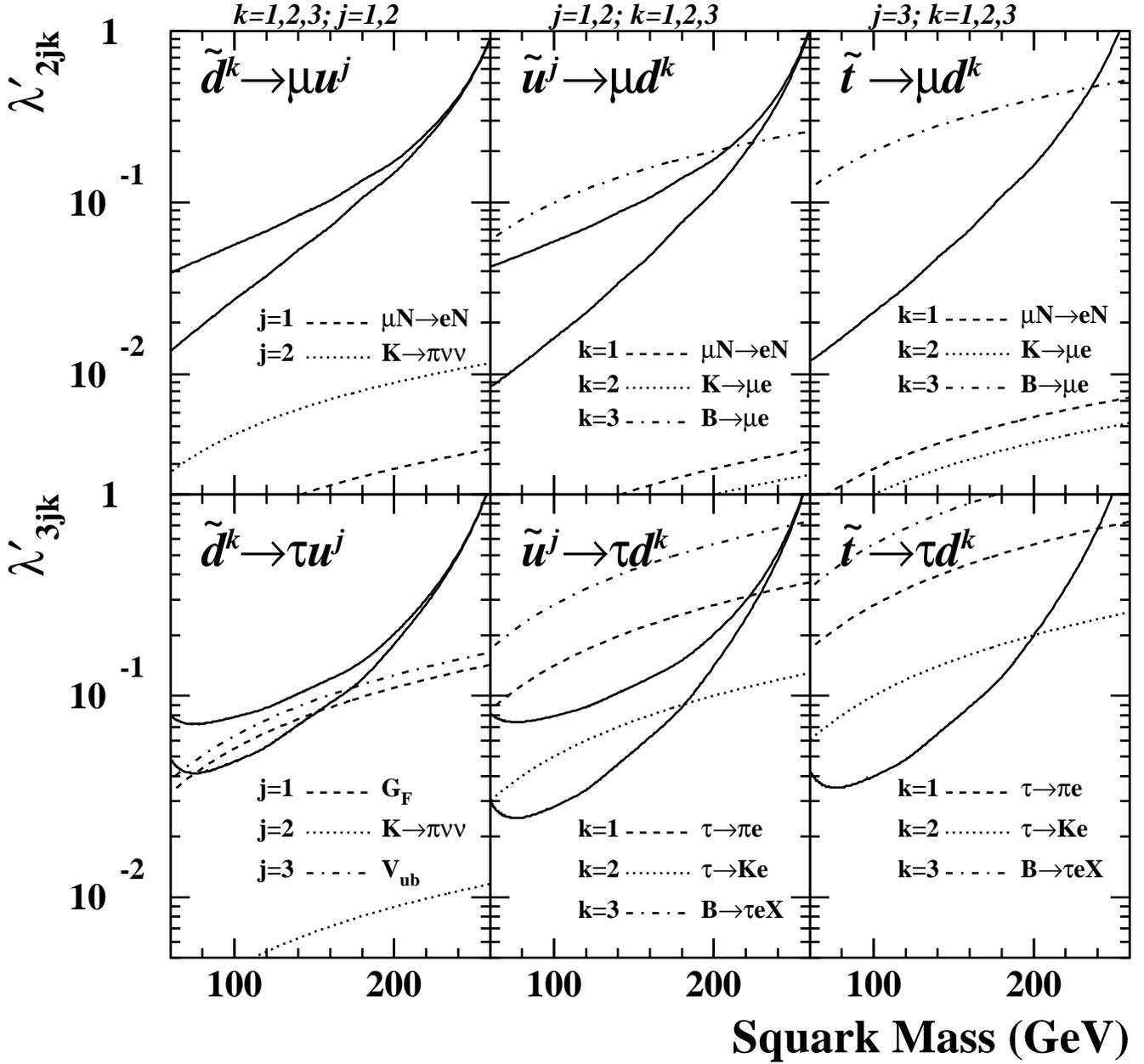,width=20cm,height=18cm}}
  \caption{
Limits on the $\rpvio$ coupling $\lambda'_{ijk}$ at $95\%$ CL for
squarks which decay to $\mu q$ ($i=2$) or $\tau q$ ($i=3$).
For ${\tilde d}$ limits, we assume that $\lambda_{11k}'=\lambda_{ijk}'$,
while for ${\tilde u}$ (including ${\tilde t}$) limits,
we assume that $\lambda_{1j1}'=\lambda_{ijk}'$.
The lower solid curves give the ZEUS limits
for squarks which decay purely via $R$-parity violation.
The upper solid curves give the ZEUS limits for the case where,
in addition, the gauge decay $\tilde{q}\to q\photino$ exists, and where
the photino is much lighter than the squark.
In the case of $\sTop$, we only consider $\rpvio$ decays and assume that the
stop mixing angle is given by $\cos^{2}\theta_t =0.5$.
The values of $j$ and $k$ for which each limit applies are indicated
above the plots.
The dashed and dotted curves show the limits from other experiments
(adapted from \protect\cite{davidson}). These limits
depend on the generation of the quark which couples to the $\mu$ or
$\tau$ as indicated.}
\label{fig:sqlimit}
\end{figure}

Coupling limits for $S_0$ and $\tilde{S}_{1/2}$
leptoquarks were converted to coupling limits on ${\tilde d}$, ${\tilde u}$,
and ${\tilde t}$ as described in section~\ref{RPVsquark}.
Figure~\ref{fig:sqlimit} shows 95\% CL limits on coupling 
{\it vs.}~mass for  $\rpvio$ squarks which decay to $\mu q$ and $\tau q$.
Here we assume the couplings
at the production vertex ($\lambda_{11k}'$ for ${\tilde d}^k$,
$\lambda_{1j1}'$ for ${\tilde u}^j$) and at the decay vertex ($\lambda_{ijk}'$)
to be equal.
The solid curves are the ZEUS limits which are given for two assumptions.
The lower curves (and the $\tilde{t}$ limits) assume that all squark
decays are $R$-parity violating. The upper curves illustrate the impact
of gauge decays on the limits. They assume that a single $R$-parity
conserving decay, namely ${\tilde q}\rightarrow q\photino$ exists and
that the photino is much lighter than the squark. Since our analysis
is not sensitive to such decays (for which the branching fraction
is given by equation \ref{eq:translqsq}) these limits are somewhat weaker.
The stop mixing angle is assumed to be $\cos^2\theta _t=0.5$.
The dashed curves are limits
from  low--energy experiments, adapted from reference \cite{davidson}.
Table~\ref{tab:smlimsum} gives lower mass limits for
$\tilde{d}$, $\tilde{u}$, and $\tilde{t}$ assuming that the couplings
at the production and decay vertices are equal to the
electromagnetic coupling ($\sqrt{4\pi\alpha}$).
As with the low--mass leptoquark case described earlier, the ZEUS limits
improve on  existing limits in cases where quark flavor change accompanies
the lepton flavor change, especially for $e\leftrightarrow\tau$ flavor changes.

\section{Conclusions}

We have searched for signatures of lepton--flavor violation
with the ZEUS detector. Hypothetical exotic particles such as
leptoquarks could induce lepton--flavor violation observable at HERA. The tight
constraints from sensitive searches for processes such as muon conversion
in titanium and rare muon and 
meson decays do not apply to all possible cases of LFV,
many of which could be seen in $ep$ collisions. Using 3.8 \ipb
of data taken at HERA during the 1993 and 1994 running periods, we have found
no candidate events for LFV.
The data permit us to constrain  specific leptoquark
coupling strengths as small as $10^{-3}\alpha$ and to exclude leptoquark
masses as large as 270 GeV (for electromagnetic coupling) with 95\%
confidence. For $\MLQ\gg \sqrt{s}$, we calculate
upper limits on the product of lepton flavor violating couplings
divided by the square of the leptoquark mass, 
$\lambda _{eq_1}\lambda_{\lepton q_2}/\MLQsq$, and directly compare these with
existing bounds from rare decays. Especially for $e\leftrightarrow\tau$
flavor changes, ZEUS has improved on existing limits for many
flavor--violating scenarios.

\section{Acknowledgments}
We thank the HERA machine group for the excellent
machine operation which made this work possible,
the DESY computing and network group for
providing the necessary data analysis environment, and
the DESY directorate for strong support and encouragement. 
We wish to thank S.~Davidson and H.~Dreiner for useful discussions.

\newpage
\section{Appendix}

We summarize here the cross section formulae \cite{buchmuller}
for processes involving 
leptoquarks which couple only to left--handed or right--handed leptons.
Process (1) can be mediated by either $s$-channel or $u$-channel
leptoquark exchange. For the $s$-channel process, $eq_1\to\lepton q_2$,
the differential cross section, for unpolarized beams, can be written as:

\begin{equation}
\frac{d^2\sigma }{dxdy}=\frac 1{32\pi xs} q_1(x,\shat)
\frac{\lambda _{eq_1}^2\lambda_{\lepton q_2}^2s^2x^2}
{(sx-\MLQsq)^2+\MLQsq\GammaLQsq}
\times\left\{\begin{array}{cc}
\frac 12 & \mbox{\rm scalar}\; LQ \\
2(1-y)^2 & \mbox{\rm vector}\; LQ,\end{array}\right.
\label{sChannel}\end{equation}
where $q_1(x,\shat)$ is the parton density\footnote{
For $s$- and $u$-channel processes, we have used $\shat$ and $-u$ 
respectively as the scale in the parton densities. If we had used
$Q^2$, the calculated cross sections would vary by less than 4\%
for initial--state $u$ and $d$ quarks and by less than 16\% for
initial--state $s$,  $c$, or $b$ quarks.
}
for the initial state quark or antiquark, $\lambda_{eq_1}$ and
$\lambda_{\lepton q_2}$
are the couplings at the production and decay vertices, and
$\GammaLQ$ is the total width of the leptoquark. The partial width
for decay into lepton $\lepton$, and quark $q_2$, is

\begin{equation}
\Gamma_{\lepton q_2}=\MLQ\lambda _{\lepton q_2}^2\times\left\{
\begin{array}{cc}
\frac1{16\pi} & {\mbox{\rm scalar}\;LQ} \\ 
\frac1{24\pi} & {\mbox{\rm vector}\;LQ,}
\end{array}\right.
\label{gamma}\end{equation}

so that the typical LQ sought here has $\GammaLQ\ll \MLQ$. In the narrow
width approximation, which holds when the variation of $q_1(x)$ is
small as $x$ is varied by $\delta x\sim\GammaLQ/\MLQ$,
integration of equation (\ref{sChannel}) leads to the formula \ref{lowmasscs},
with $B_{\lepton q_2}=\Gamma_{\lepton q_2}/\GammaLQ$.

For the $u$-channel process $eq_2\to\lepton q_1$, the differential
cross section is given by:
\begin{equation}
\frac{d^2\sigma }{dxdy}=\frac 1{32\pi xs} q_2(x,-u)
\frac{\lambda _{eq_2}^2\lambda _{lq_1}^2\,s^2\,x^2}
{\left[ sx(1-y)+\MLQsq\right] ^2}
\times\left\{\begin{array}{cc}
\frac 12 (1-y)^2 & {\mbox{\rm scalar}\; LQ} \\
2                & {\mbox{\rm vector}\; LQ.}\end{array}\right.
\label{uChannel}\end{equation}

In the limit that $\MLQsq\gg s$,
integration of equations (\ref{sChannel}) and (\ref{uChannel})
lead to equations (\ref{scs}) and (\ref{ucs}),
which are accurate to better than 10\% for $\MLQ>500$ GeV.

Note that any leptoquark will take part in both $s$- and $u$- channel
interactions. For example an $S_0^R$ leptoquark will mediate the
$s$-channel process $e^+ {\overline u}\rightarrow \mu^+ {\overline c}$
as well as the $u$-channel reaction $e^+ c\rightarrow \mu^+ u$.
\newpage
%
%

%
%
\newpage
\begin{table}
\[
\begin{tabular}{|c||c|c|c|c|c|c|c|}
\hline \rule{0mm}{8mm}
LQ species                &   $S_0^L$ &  $S_0^R$ & $\tS_0^R$ & $S_{1/2}^L$ & $S_{1/2}^R$ & $\tS_{1/2}^L$ & $S_1^L$ \\ \hline
\mbox{$\mu q$}\thinspace  &       231 &      242 &       214 &         258 &         259 &           234 &     243 \\ \hline
\mbox{$\tau q$}\thinspace &       223 &      236 &       207 &         253 &         254 &           228 &     236 \\ \hline
\hline \rule{0mm}{8mm}
LQ species                &   $V_0^L$ &  $V_0^R$ & $\tV_0^R$ & $V_{1/2}^L$ & $V_{1/2}^R$ & $\tV_{1/2}^L$ & $V_1^L$ \\ \hline
\mbox{$\mu q$}\thinspace  &       234 &      243 &       264 &         225 &         254 &           252 &     272 \\ \hline
\mbox{$\tau q$}\thinspace &       227 &      237 &       261 &         219 &         248 &           246 &     270 \\ \hline
\end{tabular}
\]
\caption{The $95\%$ confidence level 
lower limits on the LQ mass (GeV) for nominal electromagnetic coupling.
For leptoquarks which couple to $\mu$, we set $\lambda_{eq_1}^2 = \lambda_{\mu q_2}^2 = 4 \pi\alpha$.
For leptoquarks which couple to $\tau$, we set $\lambda_{eq_1}^2 = \lambda_{\tau q_2}^2 = 4 \pi\alpha$.
Limits are shown for all scalar (S) and vector (V) leptoquark species.}
\label{tab:mlimsum}
\end{table}
\newpage
\newcommand{\ubar}{\mbox{$\bar{u}$}}
\newcommand{\dbar}{\mbox{$\bar{d}$}}
\newcommand{\ebar}{\mbox{$\bar{e}$}}
\newcommand{\mubar}{\mbox{$\bar\mu$}}
\newcommand{\nubar}{\mbox{$\bar\nu$}}
\newcommand{\muN}{\mbox{$\mu N\rightarrow eN$}}
\newcommand{\Kpi}{\mbox{$K\rightarrow \pi \nubar\nu $}}
\newcommand{\Kmue}{\mbox{$K\rightarrow \mu \ebar$}}
\newcommand{\Dmue}{\mbox{$D\rightarrow \mu \ebar$}}
\newcommand{\Bmue}{\mbox{$B\rightarrow \mu \ebar$}}
\newcommand{\muegam}{\mbox{$\mu\rightarrow e\gamma$}}
\newcommand{\BlnuX}{\mbox{$B\rightarrow \lepton\nu X$}}
\newcommand{\BMueK}{\mbox{$B\rightarrow \mubar eK$}}
\newcommand{\taupie}{\mbox{$\tau \rightarrow \pi e$}}
\newcommand{\tauKe}{\mbox{$\tau \rightarrow Ke$}}
\newcommand{\BtaueX}{\mbox{$B\rightarrow \tau \ebar X$}}
\newcommand{\tauegam}{\mbox{$\tau \rightarrow e\gamma $}}
\newcommand{\Kpinunu}{\mbox{$K\rightarrow \pi \nubar\nu $}}
\begin{table}
\begin{footnotesize}\begin{displaymath}\begin{tabular}{|c||c|c|c|c|c|c|c|}
\multicolumn{8}{c}{} \\ \hline
\multicolumn{8}{|c|}{\rule{0mm}{8mm}\Large $e\leftrightarrow\mu$\hskip4cm$F=2$} \\ \hline
\rule{0mm}{8mm}&$S_0^L$        & $S_0^R$           & $\tilde{S}_0^R$     & $S_1^L$                & $V_{1/2}^L$       & $V_{1/2}^R$       & $\tilde{V}_{1/2}^L$ \\ 
$(q_1q_2)$ & $e^-u$            & $e^-u$            & $e^-d$              & $e^-(u+\sqrt2d)$     & $e^-d$            & $e^-(u+d)$        & $e^-u$            \\
           &$\nu\;\, d$        &                   &                     &$\nu\;\, (\sqrt2u+d)$ & $\nu\;\,d$        &                   & $\nu\;\,u$        \\ \hline\hline
(11)       & \muN              & \muN              & \muN                & \muN                   & \muN              & \muN              & \muN              \\ 
           & $2\times 10^{-6}$ & $2\times 10^{-6}$ & $2\times 10^{-6}$   & $5\times10^{-7}$       & $7\times 10^{-7}$ & $4\times 10^{-7}$ & $7\times 10^{-7}$ \\
           &  0.09             &  0.09             &  0.12               &  0.05                  &  0.05             &  0.03             &  0.04             \\ \hline
(12)       &  \Kpi             & \Dmue             & \Kmue               & \Kmue                  & \Kmue             & \Kmue             & \Dmue             \\ 
           & $2\times 10^{-5}$ &  0.14             & $10^{-6}$           & $6\times 10^{-7}$      & $6\times10^{-7}$  & $6\times 10^{-7}$ &  0.07             \\
           &  0.12             &  {\bf 0.12 }      &  0.14               &  0.06                  &  0.09             &  0.06             &  0.08             \\ \hline
(13)       & $V_{ub}$          &                   &  \Bmue              & $V_{ub}$               & \Bmue             & \Bmue             &                   \\ 
           &  0.004            &                   &  0.01               &  0.004                 &  0.005            &  0.005            &                   \\
           &                   &  *                &  0.15               &  0.07                  &  0.10             &  0.10             &  *                \\ \hline
(21)       & \Kpi              & \Dmue             & \Kmue               & \Kmue                  & \Kmue             & \Kmue             & \Dmue             \\ 
           & $2\times 10^{-5}$ &  0.14             & $10^{-6}$           & $6\times 10^{-7}$      & $6\times10^{-7}$  & $6\times 10^{-7}$ &  0.07             \\
           &  0.12             &  {\bf 0.12 }      &  0.14               &  0.06                  &  0.05             &  0.03             &  {\bf 0.04 }      \\ \hline
(22)       & \muegam           & \muegam           &  \muegam            & \muegam                &  \muegam          & \muegam           & \muegam           \\ 
           & $2\times 10^{-4}$ & $2\times 10^{-4}$ & $8\times 10^{-5}$   & $4\times10^{-5}$       &  0.15             & $6\times 10^{-3}$ & $6\times 10^{-3}$ \\
           &  0.24             &  0.24             &  0.20               &  0.09                  &  {\bf 0.10 }      &  0.08             &  0.13             \\ \hline
(23)       & \BlnuX            &                   & \BMueK              & \BMueK                 & \BMueK            & \BMueK            &                   \\ 
           &  0.04             &                   & $6\times 10^{-3}$   & $3\times 10^{-3}$      & $3\times 10^{-3}$ & $3\times 10^{-3}$ &                   \\
           &                   &  *                &  0.21               &  0.11                  &  0.13             &  0.13             &  *                \\ \hline
(31)       & $V_{ub}$          &                   &  \Bmue              & $V_{ub}$               & \Bmue             &  \Bmue            &                   \\ 
           &  0.004            &                   &  0.01               &  0.004                 &  0.005            &  0.005            &                   \\
           &                   &  *                &  0.16               &  0.08                  &  0.05             &  0.05             &  *                \\ \hline
(32)       & \BlnuX            &                   & \BMueK              & \BMueK                 & \BMueK            & \BMueK            &                   \\ 
           &  0.04             &                   & $6\times 10^{-3}$   & $3\times 10^{-3}$      & $3\times 10^{-3}$ & $3\times 10^{-3}$ &                   \\
           &                   &  *                &  0.26               &  0.13                  &  0.11             &  0.11             &  *                \\ \hline
(33)       &                   &                   & \muegam             & \muegam                & \muegam           & \muegam           &                   \\ 
           &                   &                   & $8\times 10^{-5}$   & $4\times 10^{-5}$      &  0.01             &  0.01             &                   \\
           &                   &  *                &  0.29               &  0.14                  &  0.15             &  0.15             &  *                \\ \hline
\end{tabular}\end{displaymath}\end{footnotesize}
\caption{The best upper bounds on $\lambda_{eq_1}\lambda_{\mu q_2}/\MLQsq$
for $F=2$ leptoquarks, in units of
$10^{-4}$ GeV$^{-2}$. Each column corresponds to a given leptoquark species
and each row to the quark flavors $q_1$ and $q_2$ which couple to $e$ and $\mu$,
the generation indices of which are specified in the first column.
The top line in each box gives the previous measurement \protect\cite{davidson}
which had obtained the strictest limit. The limit from that experiment
is given on the second line in the box and the ZEUS limit, shown on
the third line, is printed in boldface if it supersedes the previous
limit. The asterisks denote those cases where lepton flavor violation
occurs only via processes involving top.}
\label{tab:muqF2}\end{table}
\begin{table}
\begin{footnotesize}\begin{displaymath}\begin{tabular}{|c||c|c|c|c|c|c|c|}
\multicolumn{8}{c}{} \\ \hline
\multicolumn{8}{|c|}{\rule{0mm}{8mm}\Large $e\leftrightarrow\mu$\hskip4cm$F=0$} \\ \hline
\rule{0mm}{8mm}& $S_{1/2}^L$   & $S_{1/2}^R$       & $\tilde{S}_{1/2}^L$ & $V_0^L$                & $V_0^R$           & $\tilde{V}_0^R$   & $V_1^L$           \\  
$(q_1q_2)$ &$e^-\ubar$         & $e^-(\ubar+\dbar)$&$e^-\dbar$ &$e^-\dbar$ &     $e^-\dbar$       & $e^-\ubar$        & $e^-(\sqrt2\ubar+\dbar)$ \\
           &$\nu\;\,\ubar$     &                   &  $\nu\;\,\dbar$     &   $\nu\;\,\ubar$       &                   &                   &$\nu \;\,(\ubar+\sqrt2\dbar)$ \\ \hline\hline
(11)       &\muN               & \muN              & \muN                & \muN                   & \muN              & \muN              & \muN              \\ 
           &$2\times 10^{-6}$  & $7\times 10^{-7}$ & $2\times 10^{-6}$   & $7\times10^{-7}$       & $7\times 10^{-7}$ & $7\times 10^{-7}$ & $3\times 10^{-7}$ \\
           &  0.07             &  0.06             &  0.10               &  0.06                  &  0.06             &  0.04             &  0.02             \\ \hline
(12)       &  \Dmue            & \Kmue             & \Kmue               & \Kmue                  & \Kmue             & \Dmue             & \Kmue             \\ 
           &   0.14            & $10^{-6}$         & $10^{-6}$           & $6\times 10^{-7}$      & $6\times 10^{-7}$ & 0.07              & $6\times 10^{-7}$ \\
           &  {\bf 0.08 }      &  0.06             &  0.10               &  0.07                  &  0.07             & {\bf 0.06 }       &  0.03             \\ \hline
(13)       &                   &  \Bmue            &  \Bmue              & $V_{bu}$               &  \Bmue            &                   & $V_{bu}$          \\ 
           &                   &  0.01             &  0.01               &  0.002                 &  0.005            &                   &  0.002            \\
           &  *                &  0.11             &  0.11               &  0.08                  &  0.08             &  *                &  0.08             \\ \hline
(21)       &\Dmue              & \Kmue             & \Kmue               & \Kmue                  & \Kmue             & \Dmue             & \Kmue             \\ 
           & 0.14              & $10^{-6}$         & $10^{-6}$           & $6\times 10^{-7}$      & $6\times 10^{-7}$ &  0.07             & $6\times 10^{-7}$ \\
           &  0.17             &  0.12             &  0.17               &  0.07                  &  0.07             &  {\bf 0.06 }      &  0.03             \\ \hline
(22)       &\muegam            & \muegam           &                     &  \muegam               & \muegam           & \muegam           & \muegam           \\ 
           &$5\times 10^{-5}$  & $5\times 10^{-5}$ &                     &  0.07                  &  0.07             & $9\times10^{-3}$  & $5\times 10^{-3}$ \\
           &  0.24             &  0.16             &  {\bf 0.20 }        &  0.10                  &  0.10             &  0.13             &  0.05             \\ \hline
(23)       &                   & \BMueK            & \BMueK              & \BMueK                 & \BMueK            &                   & \BMueK            \\ 
           &                   & $6\times 10^{-3}$ & $6\times 10^{-3}$   & $3\times 10^{-3}$      & $3\times10^{-3}$  &                   & $3\times 10^{-3}$ \\
           &  *                &  0.21             &  0.21               &  0.13                  &  0.13             &  *                &  0.13             \\ \hline
(31)       &                   &  \Bmue            &  \Bmue              & $V_{bu}$               &  \Bmue            &                   & $V_{bu}$          \\ 
           &                   &  0.01             &  0.01               &  0.002                 &  0.005            &                   &  0.002            \\
           &  *                &  0.20             &  0.20               &  0.07                  &  0.07             &  *                &  0.07             \\ \hline
(32)       &                   & \BMueK            & \BMueK              & \BMueK                 & \BMueK            &                   & \BMueK            \\ 
           &                   & $6\times 10^{-3}$ & $6\times 10^{-3}$   & $3\times 10^{-3}$      & $3\times10^{-3}$  &                   & $3\times 10^{-3}$ \\
           &  *                &  0.26             &  0.26               &  0.11                  &  0.11             &  *                &  0.11             \\ \hline
(33)       &                   &                   &                     & \muegam                & $\muegam $        &                   & \muegam           \\ 
           &                   &                   &                     &  0.001                 &  0.001            &                   &  0.001            \\
           &  *                &  {\bf 0.29 }      &  {\bf 0.29 }        &  0.15                  &  0.15             &  *                &  0.15             \\ \hline
\end{tabular}\end{displaymath}\end{footnotesize}
\caption{The best upper bounds on $\lambda_{eq_1}\lambda_{\mu q_2}/\MLQsq$
for $F=0$ leptoquarks, in units of
$10^{-4}$ GeV$^{-2}$. Each column corresponds to a given leptoquark species
and each row to the quark flavors $q_1$ and $q_2$ which couple to $e$ and $\mu$,
the generation indices of which are specified in the first column.
The top line in each box gives the previous measurement \protect\cite{davidson}
which had obtained the strictest limit. The limit from that experiment
is given on the second line in the box and the ZEUS limit, shown on
the third line, is printed in boldface if it supersedes the previous
limit. The asterisks denote those cases where lepton flavor violation
occurs only via processes involving top.}
\label{tab:muqF0}\end{table}
\begin{table}
\begin{footnotesize}\begin{displaymath}\begin{tabular}{|c||c|c|c|c|c|c|c|}
\multicolumn{8}{c}{} \\ \hline
\multicolumn{8}{|c|}{\rule{0mm}{8mm}\Large $e\leftrightarrow\tau$\hskip4cm$F=2$} \\ \hline
\rule{0mm}{8mm}&$S_0^L$        & $S_0^R$           & $\tilde{S}_0^R$     & $S_1^L$                & $V_{1/2}^L$       & $V_{1/2}^R$       & $\tilde{V}_{1/2}^L$ \\ 
$(q_1q_2)$ & $e^-u$            & $e^-u$            & $e^-d$              & $e^-(u+\sqrt2d)$     & $e^-d$            & $e^-(u+d)$        & $e^-u$            \\
           &$\nu\;\, d$        &                   &                     &$\nu\;\, (\sqrt2u+d)$ & $\nu\;\,d$        &                   & $\nu\;\,u$        \\ \hline\hline
(11)       & $G_F$             &  \taupie          &  \taupie            & $G_F$                  &  \taupie          &  \taupie          &  \taupie          \\ 
           &  0.003            &  0.02             &  0.02               &  0.003                 &  0.01             &  0.005            &  0.01             \\
           &  0.15             &  0.15             &  0.23               &  0.09                  &  0.09             &  0.05             &  0.06             \\ \hline
(12)       &  \Kpi             &                   &  \tauKe             &  \Kpi                  &  \Kpi             &  \tauKe           &                   \\ 
           & $2\times 10^{-5}$ &                   &  0.05               & $2\times 10^{-5}$      & $10^{-5}$         &  0.03             &                   \\
           &  0.20             &  {\bf 0.20 }      &  0.27               &  0.11                  &  0.19             &  0.13             &  {\bf 0.16 }      \\ \hline
(13)       & $V_{bu}$          &                   &  \BtaueX            & $V_{bu}$               & \BtaueX           &  \BtaueX          &                   \\ 
           &  0.004            &                   &  0.08               &  0.004                 &  0.04             &  0.04             &                   \\
           &                   &  *                &  0.28               &  0.14                  &  0.23             &  0.23             &   *               \\ \hline
(21)       &  \Kpi             &                   &  \tauKe             &  \Kpi                  &  \Kpi             &  \tauKe           &                   \\ 
           & $2\times 10^{-5}$ &                   &  0.05               & $2\times 10^{-5}$      & $10^{-5}$         &  0.03             &                   \\
           &  0.22             &  {\bf 0.22 }      &  0.31               &  0.12                  &  0.09             &  0.05             &  {\bf 0.06 }      \\ \hline
(22)       & \tauegam          & \tauegam          & \tauegam            & \tauegam               &                   &                   &                   \\ 
           &  0.5              &  0.5              &  0.3                &  0.1                   &                   &                   &                   \\
           &  0.60             &  0.60             &  0.48               &  0.22                  &  {\bf 0.25 }      &  {\bf 0.19 }      &  {\bf 0.31 }      \\ \hline
(23)       & \BlnuX            &                   &  \BtaueX            & \BlnuX                 &  \BtaueX          &  \BtaueX          &                   \\ 
           &  0.04             &                   &  0.08               &  0.04                  &  0.04             &  0.04             &                   \\
           &                   &  *                &  0.50               &  0.25                  &  0.33             &  0.33             &   *               \\ \hline
(31)       & \BlnuX            &                   &  \BtaueX            & \BlnuX                 &  \BtaueX          &  \BtaueX          &                   \\ 
           &  0.04             &                   &  0.08               &  0.04                  &  0.04             &  0.04             &                   \\
           &                   &  *                &  0.34               &  0.17                  &  0.10             &  0.10             &   *               \\ \hline
(32)       & \BlnuX            &                   &  \BtaueX            & \BlnuX                 &  \BtaueX          &  \BtaueX          &                   \\ 
           &  0.04             &                   &  0.08               &  0.04                  &  0.04             &  0.04             &                   \\
           &                   &  *                &  0.65               &  0.32                  &  0.26             &  0.26             &   *               \\ \hline
(33)       &                   &                   & \tauegam            & \tauegam               &                   &                   &                   \\ 
           &                   &                   &  0.3                &  0.1                   &                   &                   &                   \\
           &                   &  *                &  0.72               &  0.36                  &  {\bf 0.38 }      &  {\bf 0.38 }      &   *               \\ \hline
\end{tabular}\end{displaymath}\end{footnotesize}
\caption{The best upper bounds on $\lambda_{eq_1}\lambda_{\tau q_2}/\MLQsq$
for $F=2$ leptoquarks, in units of
$10^{-4}$ GeV$^{-2}$. Each column corresponds to a given leptoquark species
and each row to the quark flavors $q_1$ and $q_2$ which couple to $e$ and $\tau$,
the generation indices of which are specified in the first column.
The top line in each box gives the previous measurement \protect\cite{davidson}
which had obtained the strictest limit. The limit from that experiment
is given on the second line in the box and the ZEUS limit, shown on
the third line, is printed in boldface if it supersedes the previous
limit. The asterisks denote those cases where lepton flavor violation
occurs only via processes involving top.}
\label{tab:tauqF2}\end{table}
\begin{table}
\begin{footnotesize}\begin{displaymath}\begin{tabular}{|c||c|c|c|c|c|c|c|}
\multicolumn{8}{c}{} \\ \hline
\multicolumn{8}{|c|}{\rule{0mm}{8mm}\Large $e\leftrightarrow\tau$\hskip4cm$F=0$} \\ \hline
\rule{0mm}{8mm}& $S_{1/2}^L$   & $S_{1/2}^R$       & $\tilde{S}_{1/2}^L$ & $V_0^L$                & $V_0^R$           & $\tilde{V}_0^R$   & $V_1^L$           \\  
$(q_1q_2)$ &$e^-\ubar$         &$e^-(\ubar+\dbar)$ & $e^-\dbar$          & $e^-\dbar$             & $e^-\dbar$        & $e^-\ubar$        & $e^-(\sqrt2\ubar+\dbar)$ \\
           &$\nu\;\,\ubar$     &                   & $\nu\;\,\dbar$      & $\nu\;\,\ubar$         &                   &                   & $\nu \;\,(\ubar+\sqrt2\dbar)$ \\ \hline\hline
(11)       &  \taupie          &  \taupie          &  \taupie            & $G_F$                  &  \taupie          &  \taupie          & $G_F$             \\ 
           &  0.02             &  0.01             &  0.02               &  0.002                 &  0.01             &  0.01             &  0.002            \\
           &  0.11             &  0.09             &  0.18               &  0.11                  &  0.11             &  0.07             &  0.04             \\ \hline
(12)       &                   &  \tauKe           &  \Kpi               &  \tauKe                &  \tauKe           &                   & \Kpinunu          \\ 
           &                   &  0.05             & $2\times 10^{-5}$   &  0.03                  &  0.03             &                   & $5\times 10^{-6}$ \\
           &  {\bf 0.12 }      &  0.10             &  0.18               &  0.15                  &  0.15             &  {\bf 0.10 }      &  0.05             \\ \hline
(13)       &                   &  \BtaueX          &  \BtaueX            & \BlnuX                 &  \BtaueX          &                   & \BlnuX            \\ 
           &                   &  0.08             &  0.08               &  0.02                  &  0.04             &                   &  0.02             \\
           &  *                &  0.18             &  0.18               &  0.16                  &  0.16             &  *                &  0.16             \\ \hline
(21)       &                   &  \tauKe           &  \Kpi               &  \tauKe                &  \tauKe           &                   & \Kpinunu          \\ 
           &                   &  0.05             & $2\times 10^{-5}$   &  0.03                  &  0.03             &                   & $5\times 10^{-6}$ \\
           &  {\bf 0.34 }      &  0.26             &  0.39               &  0.14                  &  0.14             &  {\bf 0.10 }      &  0.05             \\ \hline
(22)       & \tauegam          & \tauegam          &                     &                        &                   &                   &                   \\ 
           &  0.2              &  0.2              &                     &                        &                   &                   &                   \\
           &  0.60             &  0.37             &  {\bf 0.48 }        &  {\bf 0.25 }           &  {\bf 0.25 }      &  {\bf 0.31 }      &  {\bf 0.13 }      \\ \hline
(23)       &                   &  \BtaueX          &  \BtaueX            & \BlnuX                 &  \BtaueX          &                   & \BlnuX            \\ 
           &                   &  0.08             &  0.08               &  0.02                  &  0.04             &                   &  0.02             \\
           &  *                &  0.50             &  0.50               &  0.33                  &  0.33             &  *                &  0.33             \\ \hline
(31)       &                   &  \BtaueX          &  \BtaueX            & $V_{bu}$               &  \BtaueX          &                   & $V_{bu}$          \\ 
           &                   &  0.08             &  0.08               &  0.002                 &  0.04             &                   &  0.002            \\
           &  *                &  0.47             &  0.47               &  0.15                  &  0.15             &  *                &  0.15             \\ \hline
(32)       &                   &  \BtaueX          &  \BtaueX            & \BlnuX                 &  \BtaueX          &                   & \BlnuX            \\ 
           &                   &  0.08             &  0.08               &  0.02                  &  0.04             &                   &  0.02             \\
           &  *                &  0.65             &  0.65               &  0.26                  &  0.26             &  *                &  0.26             \\ \hline
(33)       &                   &                   &                     &  \tauegam              &  \tauegam         &                   &                   \\ 
           &                   &                   &                     &   3.4                  &  3.4              &                   &                   \\
           &  *                &  {\bf 0.72 }      &  {\bf 0.72 }        &  {\bf 0.38 }           &  {\bf 0.38 }      &  *                & {\bf 0.38}        \\ \hline
\end{tabular}\end{displaymath}\end{footnotesize}
\caption{The best upper bounds on $\lambda_{eq_1}\lambda_{\tau q_2}/\MLQsq$
for $F=0$ leptoquarks, in units of
$10^{-4}$ GeV$^{-2}$. Each column corresponds to a given leptoquark species
and each row to the quark flavors $q_1$ and $q_2$ which couple to $e$ and $\tau$,
the generation indices of which are specified in the first column.
The top line in each box gives the previous measurement \protect\cite{davidson}
which had obtained the strictest limit. The limit from that experiment
is given on the second line in the box and the ZEUS limit, shown on
the third line, is printed in boldface if it supersedes the previous
limit. The asterisks denote those cases where lepton flavor violation
occurs only via processes involving top.}
\label{tab:tauqF0}\end{table}
\newpage
%
\begin{table}
\begin{tabular}{|c|c|c|c|c|c|c|}
\hline  \rule{0mm}{8mm}
&
$\tilde{d}\rightarrow\mu q$ &
$\tilde{u}\rightarrow\mu q$ &
$\tilde{t}\rightarrow\mu q$ &
$\tilde{d}\rightarrow\tau q$ &
$\tilde{u}\rightarrow\tau q$ &
$\tilde{t}\rightarrow\tau q$ \\ \hline \hline
50\% gauge decays & 217 & 223 &  -  & 209 & 216 &  -  \\ \hline
  no gauge decays & 231 & 234 & 223 & 223 & 228 & 216 \\ \hline
\end{tabular}
\caption{95\% CL mass limits (GeV) for squarks with $\rpvio$-couplings, 
of electromagnetic strength ($\lambda_{11k}^2=\lambda_{ijk}^2=4\pi\alpha=4\pi/128$).
The first line gives the mass limits for the case where the
total branching fraction for gauge decays is 50\%, the second
line gives the mass limits for a squark which always has $\rpvio$ decays.
The limits which assume 50\% gauge decays are weaker
because we did not search for the gauge decays.
The mixing angle of the stop is assumed to be $\cos^2\theta_t=0.5$.
}
\label{tab:smlimsum}
\end{table}
\clearpage
\begin{figure}
  \begin{center}
  \begin{picture}(450,150)(0,0)
    \ArrowLine(15,15)(50,75)
    \ArrowLine(100,75)(135,15)
    \ArrowLine(15,135)(50,75)
    \ArrowLine(100,75)(135,135)
    \ZigZag(50,75)(100,75){5}{5}
    \Vertex(50,75){2.5}
    \Vertex(100,75){2.5}
    \Text(40,35)[lb]{\Large$q_1$}
    \Text(110,35)[rb]{\Large$q_2$}
    \Text(40,115)[lt]{\Large$e$}
    \Text(110,115)[rt]{\Large$\lepton$}
    \Text(40,75)[r]{\Large$\lambda_{eq_1}$}
    \Text(110,75)[l]{\Large$\lambda_{\lepton q_2}$}
    \Text(75,0)[t]{\Large\bf a)}
    \ArrowLine(165,15)(225,50)
    \ArrowLine(225,50)(285,15)
    \ArrowLine(165,135)(225,100)
    \ArrowLine(225,100)(285,135)
    \ZigZag(225,50)(225,100){5}{5}
    \Vertex(225,50){2.5}
    \Vertex(225,100){2.5}
    \Text(190,40)[b]{\Large$q_2$}
    \Text(260,40)[b]{\Large$\lepton$}
    \Text(190,110)[t]{\Large$e$}
    \Text(260,110)[t]{\Large$q_1$}
    \Text(225,20)[b]{\Large$\lambda_{\lepton q_2}$}
    \Text(225,130)[t]{\Large$\lambda_{eq_1}$}%
    \Text(225,0)[t]{\Large\bf b)}
    \ArrowLine(315,15)(375,50)
    \ArrowLine(375,50)(435,15)
    \ArrowLine(315,135)(375,100)
    \ArrowLine(375,100)(435,135)
    \ZigZag(375,50)(375,100){5}{5}
    \Vertex(375,50){2.5}
    \Vertex(375,100){2.5}
    \Text(340,40)[b]{\Large$q_1$}
    \Text(410,40)[b]{\Large$q_2$}
    \Text(340,110)[t]{\Large$e$}
    \Text(410,110)[t]{\Large$\lepton$}
    \Text(375,20)[b]{\Large$\lambda_{q_1q_2}$}
    \Text(375,130)[t]{\Large$\lambda_{e\lepton}$}
    \Text(375,0)[t]{\Large\bf c)}
  \end{picture}
  \vskip8.mm
\caption{
The (a) $s$--, (b) $u$--, and (c) $t$--channel Feynman diagrams for LFV.
For the $s$--channel and $u$--channel diagrams,
we denote the couplings as $\lambda_{\lepton q}$, where the indices
refer to the lepton and quark flavors.}
 \label{fig:stugraph}
\end{center}\end{figure}
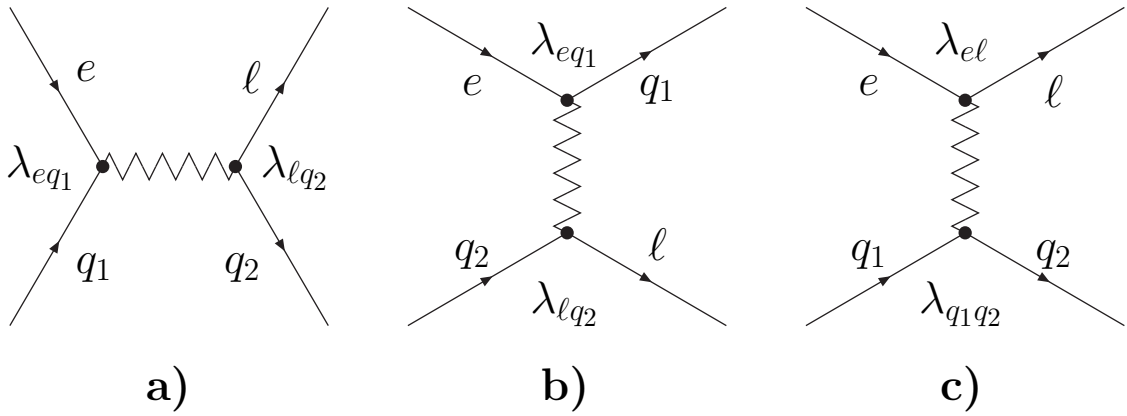
\begin{figure}
  \centerline{\psfig{figure=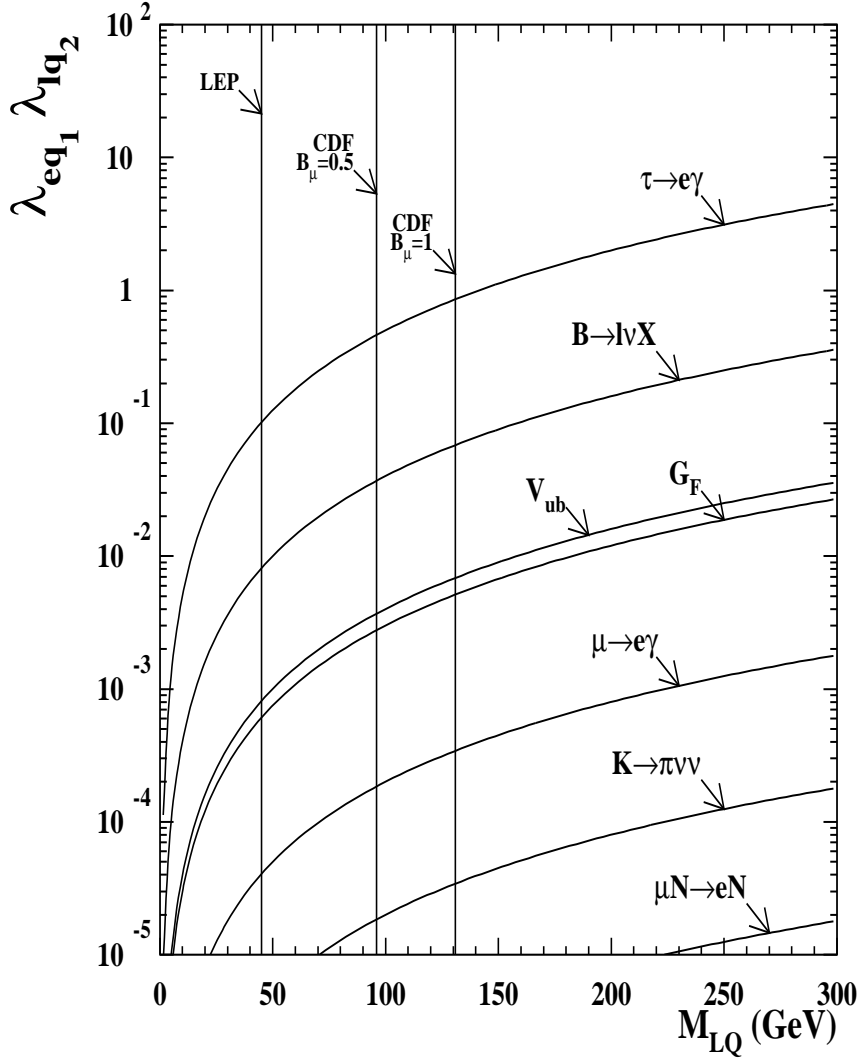,width=12cm,height=15cm}}
  \caption{
Existing 95\% CL limits \protect\cite{davidson}
 on the product of the couplings $\lambda_{eq_1}\lambda_{\lepton q_2}$
{\it vs.}~leptoquark mass $\MLQ$ for a
$S_0^L$ leptoquark mediating $e\leftrightarrow\mu$ and $e\leftrightarrow\tau$ transitions
accompanied by various quark flavor changes. Each limit curve excludes
the region above it. The vertical lines indicate lower limits of allowed
leptoquark mass from LEP \protect\cite{lep} and the
Tevatron \protect\cite{tevatron}. The
Tevatron limits apply to any scalar leptoquark which couples to $\mu$
and depend on the branching fractions $B_\mu$ to the $\mu q$ final state.
The limits are shown for $B_\mu$ equal to 0.5 and 1.}
  \label{fig:illulimit}
\end{figure}
\begin{figure}
  \begin{center}
  \begin{picture}(450,300)(0,0)
    \ArrowLine(15,285)(50,225)
    \ArrowLine(15,165)(50,225)
    \Vertex(50,225){2.5}
    \Text(40,265)[lt]{\Large${\overline e}$}
    \Text(40,225)[r]{\Large$\lambda_{1j1}'$}
    \Text(40,185)[lb]{\Large$d$}
    \DashLine(50,225)(120,225){5}
    \Text(85,230)[b]{\Large$\tilde u^j$}
    \Text(85,150)[t]{\Large\bf a)}
    \ArrowLine(120,225)(155,285)
    \ArrowLine(120,225)(155,165)
    \Vertex(120,225){2.5}
    \Text(130,265)[rt]{\Large${\overline\lepton}^i$}
    \Text(130,225)[l]{\Large$\lambda_{ijk}'$}
    \Text(130,185)[rb]{\Large$d^k$}
%
    \ArrowLine(280,285)(315,225)
    \ArrowLine(280,165)(315,225)
    \Vertex(315,225){2.5}
    \Text(305,265)[lt]{\Large$e$}
    \Text(305,225)[r]{\Large$\lambda_{11k}'$}
    \Text(305,185)[lb]{\Large$u$}
    \DashLine(315,225)(385,225){5}
    \Text(350,230)[b]{\Large$\tilde d^k$}
    \Text(350,150)[t]{\Large\bf b)}
    \ArrowLine(385,225)(420,165)
    \ArrowLine(385,225)(420,285)
    \Vertex(385,225){2.5}
    \Text(395,265)[rt]{\Large$\lepton^i$}
    \Text(395,225)[l]{\Large$\lambda_{ijk}'$}
    \Text(395,185)[rb]{\Large$u^j$}
%
    \ArrowLine(145,135)(180,75)
    \ArrowLine(145,15)(180,75)
    \Vertex(180,75){2.5}
    \Text(170,115)[lt]{\Large$e$}
    \Text(170,75)[r]{\Large$\lambda_{1j1}'$}
    \Text(170,35)[lb]{\Large$d$}
    \DashLine(190,75)(260,75){5}
    \Text(225,80)[b]{\Large$\tilde u^j$}
    \Text(225,0)[t]{\Large\bf c)}
    \Photon(260,75)(295,135){5}{4.5}
    \ArrowLine(260,75)(295,15)
    \Vertex(260,75){2.5}
    \Text(270,115)[rt]{\Large$\photino$}
    \Text(270,75)[l]{\Large$\sqrt{8\pi\alpha} e_{\tilde u}$}
    \Text(270,35)[rb]{\Large$u^j$}
  \end{picture}
  \vskip8.mm
\caption{
$R_P$ violating single squark production in $ep$ collisions.
Diagrams a) and b)
show  production of ${\tilde u}$ and ${\tilde d}$ squarks with
leptoquark--like $\rpvio$ decays, where $\lepton^i$ denotes the
final--state charged lepton of generation $i$. The indices $j$ and $k$
denote the generations of up--type and down--type (s)quarks respectively.
Diagram c) shows ${\tilde u}$ production with an $R_P$--conserving decay.}
  \label{susy:graph}
  \end{center}
\end{figure}
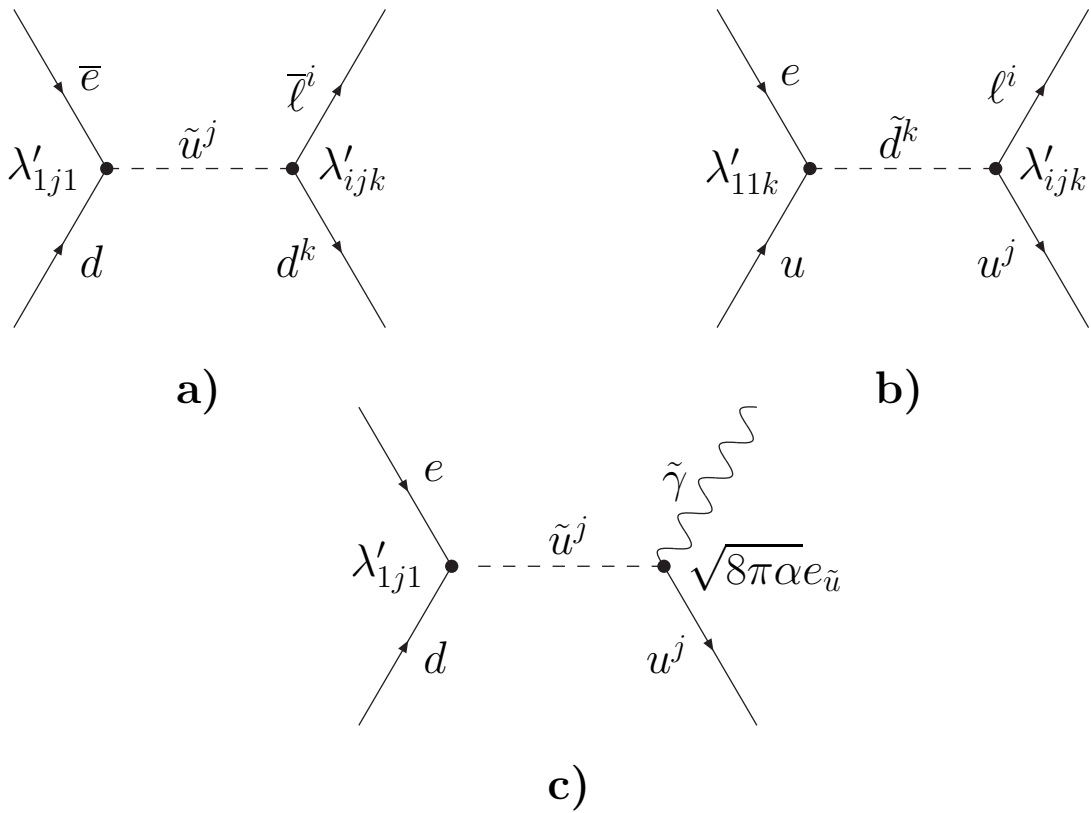
\begin{figure}
\centerline{\psfig{figure=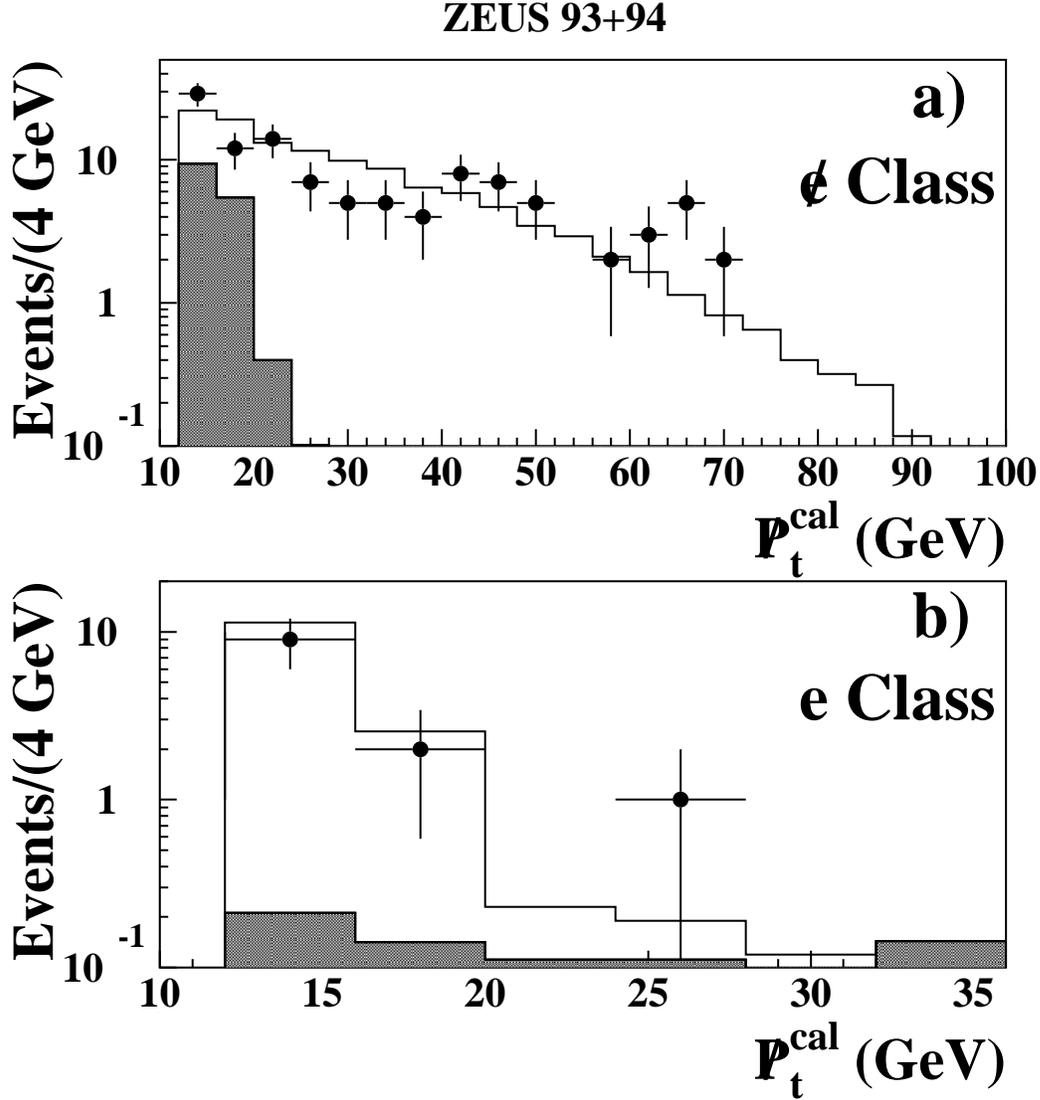,width=15cm,height=15cm}}
  \caption{
Net transverse momentum ($\ptmiss$) distribution of events in the class
$\noe$ sample (a) and the class $e$ sample (b).
The points represent the data.
The solid lines show the Monte Carlo prediction, which includes
CC DIS, NC DIS, resolved and direct photoproduction
and $\gamma\gamma$ interactions.
In the top plot, the shaded region shows the Monte
Carlo prediction for all processes except CC DIS.
In the bottom plot, the shaded region shows the Monte
Carlo prediction for all processes except NC DIS.
  }
  \label{fig:ptclass12}
\end{figure}
\end{document}